\documentstyle[prl,aps,twocolumn]{revtex}
\large  

\input epsf
\begin{document}
\draft \twocolumn[\hsize\textwidth
\columnwidth\hsize\csname
@twocolumnfalse\endcsname
\title{Ground-state properties of one-dimensional matter and quantum 
dissociation of a Luttinger liquid}
\author{Eugene B. Kolomeisky$^{*}$, Xiaoya Qi$^{\dagger}$, 
and Michael Timmins$^\#$} 

\address {Department of Physics, University of Virginia, P. O. Box 400714,
Charlottesville, VA 22904-4714}
\maketitle
\begin{abstract}
Motivated by emerging experimental possibilities to confine atoms and 
molecules in quasi-one-dimensional geometries, we analyze ground-state 
properties of strictly one-dimensional {\it molecular} matter comprised of
identical particles of mass $m$.  Such a class of systems can be described by 
an additive two-body potential whose functional form is common to all  
substances which only differ in the energy $\epsilon$ and range $l$ scales of 
the potential.  With this choice De Boer's quantum theorem of corresponding 
states holds thus implying that ground-state properties expressed in 
appropriate reduced form are only determined by the dimensionless 
parameter $\lambda_{0}^{2} \sim \hbar^{2}/ml^{2}\epsilon$ measuring the 
strength of zero-point motion in the system.  The presence of a minimum in the 
two-body interaction potential leads to a many-body bound state which is a 
Luttinger liquid stable for a not very large $\lambda_{0}$.  As $\lambda_{0}$ 
increases, the asymmetry of the two-body potential causes quantum expansion, 
softening, and eventual evaporation of the Luttinger liquid into a gas phase.  
Selecting the pair interaction potential in the Morse form we analytically 
compute the properties of the Luttinger liquid and its range of existence.  
We find that as $\lambda_{0}$ increases, the system first undergoes a 
discontinuous evaporation transition into a {\it diatomic} gas followed by a 
continuous dissociation transition into a monoatomic gas.  Two-body potentials 
of molecular systems can be successfully fitted into the Morse form thus 
allowing determination of the quantum parameter $\lambda_{0}$ and the state of 
matter of substance in question.  In particular we find that 
spin-polarized isotopes of hydrogen and $^{3}He$ are monoatomic gases, 
$^{4}He$ is a diatomic gas, while molecular hydrogen and heavier substances 
are Luttinger liquids.  We also investigate the effect of finite pressure on 
the properties of the liquid and monoatomic gas phases.  In particular we 
estimate a pressure at which molecular hydrogen undergoes an inverse Peierls 
transition into a metallic state which is a one-dimensional analog of the 
transition predicted by Wigner and Huntington in 1935.       
        
\end{abstract}
\vspace{2mm} 
]

\narrowtext

\section{Introduction}

Prediction of the ground-state properties of a condensed many-body system of 
identical particles starting from microscopic two-body interactions is only 
meaningful if the outer electronic shells of underlying atoms or molecules 
in bound phase are not very different from their free state counterparts 
\cite{Ashcroft&Mermin}.  If the two-particle potential has 
a functional form common to a family of substances (for example, of the 
Lennard-Jones type), then the properties of all the members of the family can 
be related.  This conclusion pioneered by De Boer and collaborators, commonly 
referred as {\it the quantum theorem of corresponding states}, was
originally applied to predict the properties of $^{3}He$ \cite{deBoer} 
{\it before} it had become experimentally available.  Later Anderson and Palmer
\cite{Anderson&Palmer} and Clark and Chao \cite{Clark&Chao} used the same 
approach to estimate the properties of zero-temperature nuclear and 
neutron-star matter from those of laboratory substances.   

The goal of this work is to conduct a similar program in a strictly 
one-dimensional case.  There are several reasons, both of fundamental and 
practical nature, why it is important to understand this problem. 

First, it is well-known that for ordinary substances zero-point 
motion is of crucial importance only for the lightest elements such as helium 
isotopes as well as spin-polarized isotopes of hydrogen.  The ground state of 
all heavier elements (and molecular hydrogen) is crystalline and to large 
extent classical.  The one-dimensional case is qualitatively different: 
regardless of the particle mass zero-point motion destroys the long-range 
crystalline order - a situation that is closely analogous to the destruction 
of the long-range order by thermal fluctuations in classical two-dimensional 
systems of continuous symmetry \cite{lro}.  As a result, the only possible 
many-body bound state in one dimension is a harmonic or Luttinger liquid - a 
uniform density condensed phase with algebraically decaying density 
correlations \cite{Luttinger}.  This decay, characterized by a nonuniversal 
exponent, is slower than the exponential fall-off of density correlations in 
conventional fluids.  As the degree of zero-point motion increases, the 
many-body bound state can disappear through a transition that has no analog in 
the three-dimensional world:  since the Luttinger liquid phase is more 
correlated than conventional fluids and less correlated than standard 
crystals its dissociation combines qualitative features of both laboratory 
melting and evaporation at once.  

Second, in one dimension the difference between fermions and bosons is not very
significant as we cannot go from one configuration to another with exchanged 
particles without bringing the particles in contact at some intermediate 
step.  Then short-distance repulsion among bosons has the same effect on
density correlations as the Pauli principle for fermions.  On the other hand, 
zero-temperature properties of three-dimensional matter with an
interaction pair potential of the Lennard-Jones form are sensitive to
the statistics of the underlying particles \cite{Nosanow}. 

Since experimental discovery of carbon nanotubes in 1991 \cite{Iijima} 
studying the properties of one-dimensional systems became especially 
important.  In addition to their unique transport, mechanical and 
chemical properties \cite{nanotubes1}, bundles of carbon nanotubes can play a 
role of one-dimensional hosts for foreign atoms that can find themselves 
bound in the interstitial channels or inside the tubes \cite{nanotubes2}. One 
of the interesting potential applications of these systems includes storage 
devices for molecular hydrogen in fuel cells \cite{storage}.   

Recently the quasi-one-dimensional regime has been also realized for 
Bose-condensates of alkali vapors both for repulsive \cite{repbose} and 
attractive interactions \cite{attrbose}.  These systems which are relevant 
for atom interferometry \cite{Kasevich} have an additional flexibility
as the strength and sign of two-body interactions can be magnetically tuned.

In both of these experimental examples the basic model is a zero-temperature 
one-dimensional many-body system of fermions or bosons with pairwise 
interactions. There were several attempts in the past to study this problem:

(i)  Diffusion Monte Carlo studies predicted that at zero temperature 
both one-dimensional $^{4}He$ \cite{dmc1,dmc2} and molecular hydrogen 
\cite{dmc3} form weakly-bound liquids. 

(ii)  These conclusions were supported by variational studies 
based on the Jastrow-Feenberg wave function \cite{krot} where additionally it 
was argued that in one dimension the many-body bound state exists only for 
those systems which have a dimer, i. e. a two-body bound state.  One of the 
findings common to Refs.\cite{dmc2,dmc3,krot} is the prediction of a 
high-density liquid-solid phase transition in which a standing density wave 
sets in. We note however that the existence of such a one-dimensional solid 
contradicts the quantum version of the Mermin-Wagner-Hohenberg theorem 
\cite{lro}.   

(iii) A direct variational treatment based on a Gaussian wave function was 
performed in Ref.\cite{direct} where it was assumed that the particles form a
one-dimensional chain with the Morse potential interaction \cite{Morse} 
between nearest neighbors.  Although it was shown that the chain remained 
stable for not very strong quantum fluctuations, the accuracy of the method, 
the nature of the condensed phase, the role of dimerization, and the 
implications for real systems were not addressed. 

In this paper the problem of the ground-state properties of a one-dimensional 
many-body system is re-examined for the case when the two-body interparticle 
interaction can be approximated by the Morse potential \cite{Morse}.  Since it
involves three parameters, the Morse potential is more flexible in 
describing real systems as compared to the two-parameter Lennard-Jones 
potential.  At the same time the quantum theorem of the corresponding states
\cite{deBoer} still holds in this case.  In addition the problem of the Morse
dimer is exactly solvable \cite{Morse};  below we will also show that 
analytical progress is possible in the many-body case, and the accuracy of
our results can be assessed.  
\begin{figure}[htbp]
\epsfxsize=3.6in
\vspace*{-0.3cm}
\hspace*{-0.5cm}
\epsfbox{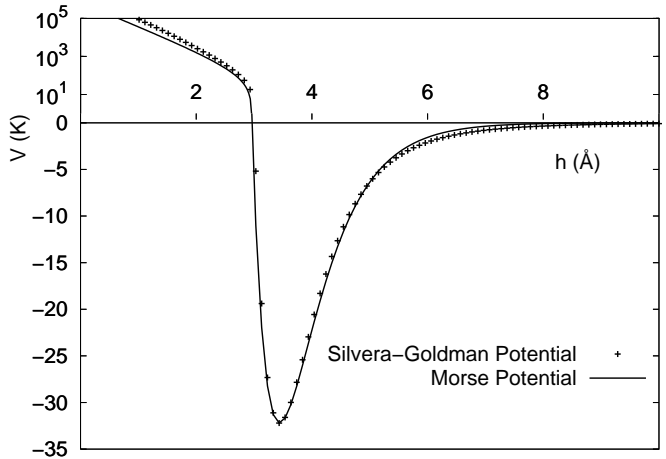}
\vspace*{0.1cm}
\caption{The Silvera-Goldman interaction potential between two hydrogen 
molecules (crosses) and its approximation by the Morse potential (solid 
curve).  To demonstrate the strength of the overlap repulsion, 
logarithmic scale is used on the positive energy axis.  The details 
of the fit are explained in the text.}
\end{figure}
The organization of the paper is as follows.  In Section II we set up the
problem in general terms with the quantum theorem of the corresponding
states \cite{deBoer} as a guide for possible outcomes.  The parameters of the 
Luttinger liquid are computed in Sections III-V.   In Section III we describe 
general properties of the Luttinger liquid and outline the main idea of the 
calculation.  Zero-pressure analysis is carried out in Section IV.  The main 
tool here is a combination of variational and renormalization-group 
treatments.  As a by-product we also solve the problem of quantum Brownian 
motion in the Morse potential, and show that it exhibits a 
localization-delocalization transition.  Finite-pressure variational analysis 
is conducted in Section V.  Section VI is dedicated to the discussion and 
applications of our results to various molecular substances.  In particular 
we estimate a pressure at which one-dimensional molecular hydrogen undergoes 
a transition into a metallic state.  

\section{Formulation of the problem}

Our starting point is the Euclidian action for $N$ identical fermions or 
bosons of mass $m$ with pairwise interactions
\begin{equation} 
\label{microaction}
S =  \int d\tau [{m \over 2} \sum \limits _{i=1}^{N}({dx_{i}\over d\tau})^{2} 
+ \sum \limits _{i<j}^{N} V(x_{i} - x_{j})],
\end{equation}
where $\tau$ is the imaginary time variable, $x_{i}$ are particle positions, 
and $V(h)$ is the pair interaction potential.  The properties of the function 
$V(h)$ can be summarized as follows.  At large separation $h$ interparticle 
interaction is dominated by weak rapidly decaying Van der Waals attraction, 
while at short distances there is a very strong overlap repulsion 
\cite{Ashcroft&Mermin}.  As a result, the pair potential $V(h)$ has an 
asymmetric minimum at some intermediate $h$.  As a typical example Fig.1 
shows the semi-empirical Silvera-Goldman potential \cite{SG} between
two $H_{2}$ molecules which is extensively used in computations of the 
properties of molecular hydrogen. 

Assume that the pair potential has the form 
\begin{equation}
\label{potential}
V(h) = \epsilon U[(h - H_{0})/l],
\end{equation}
where $\epsilon$ is the energy scale of the potential, $l$ is the potential 
range, $H_{0}$ is a length scale, and $U(y)$ is a function common to a family 
of substances.  Introducing dimensionless position 
and time variables, $q_{i} = x_{i}/l$, $t = \epsilon\tau/\hbar$, transforms 
the reduced action $S/\hbar$ into 
\begin{equation} 
\label{reducedmicroaction}
{S\over \hbar} =  \int dt [{m \epsilon l^{2} \over 2\hbar^{2}} \sum \limits _{i=1}^{N}({dq_{i}\over dt})^{2} 
+ \sum \limits _{i<j}^{N} U(q_{i} - q_{j} - Q_{0})],
\end{equation}
where $Q_{0} = H_{0}/l$.  The quantum theorem of the corresponding states
\cite{deBoer} then directly follows from representation 
(\ref{reducedmicroaction}):  the energy {\it per particle} $E^{*}$ measured in 
units of $\epsilon$ is only determined by the dimensionless parameter
\begin{equation} 
\label{lambda_{0}}
\lambda_{0} = {\hbar\over \pi l (2m\epsilon)^{1/2}}   ,
\end{equation}  
the form of the function $U$ in (\ref{potential}), and the particle statistics:
 \begin{equation}
\label{qtcs}
E^{*} = E^{*}(\lambda_{0}, statistics)
\end{equation}
Similar statements can be made about the reduced {\it relative} equilibrium 
length per particle $Q - Q_{0}$ ($Q$ is the one-dimensional version of volume 
per particle measured in units of $l$), and other quantities of interest.  
Apart from numerical factors (introduced for convenience), the quantum 
parameter $\lambda_{0}$ (\ref{lambda_{0}}) is identical to 
De Boer's number \cite{deBoer}: its square is proportional to the 
ratio of the zero-point energy of particle of mass $m$ localized within
range $l$ to the typical potential energy $\epsilon$.  Therefore as 
$\lambda_{0}$ increases away from its classical value $\lambda_{0} = 0$, the 
strength of zero-point motion increases.   
 
In the presence of several competing phases the function $E^{*}$ in
(\ref{qtcs}) (and other properties) will have several branches; the branch 
with lowest $E^{*}$ singles out the ground-state of the system.  When two 
different branches 
cross, the ground-state changes via a first-order phase transition.  
Each branch of $E^{*}(\lambda_{0})$ is an analytical function of its argument 
except possibly at isolated points where critical phenomena take place.  One 
obvious branch of $E^{*}$ corresponds to a monoatomic gas 
which must become the ground state of the system at sufficiently large 
$\lambda_{0}$.  Then all the particles are infinitely far apart from each 
other, and thus $E^{*}_{mono}(\lambda_{0}) = 0$ which we select to be the
reference point for the energy.

In what follows we select the pair interaction potential in the Morse form 
\cite{Morse}:
\begin{eqnarray}
\label{Morse}  
V(h)& =& -A e^{-h/l} + B e^{-2h/l}\nonumber\\ 
&\equiv& \epsilon[-2e^{-(h - H_{0})/l} +  e^{-2(h - H_{0})/l}],
\end{eqnarray}
where $A$ and $B$ are the amplitudes of the attractive and repulsive parts of 
the potential, respectively, $H_{0} = l \ln(2B/A)$ is the location of the 
minimum of (\ref{Morse}), while $\epsilon = V(H_{0}) = A^{2}/4B$ is the depth 
of the potential well.  The second representation of (\ref{Morse}) shows 
explicitly that the Morse potential conforms to the general form 
(\ref{potential}).  It is physically reasonable to require that the zero of 
(\ref{Morse}) is located at positive $h$ which implies $B > A$.

Similar to the applications of the Lennard-Jones potential to laboratory 
molecular systems \cite{Ashcroft&Mermin}, Eq.(\ref{Morse}) should not be 
taken too literally as really describing two-particle interactions.  The only 
reason behind our choice (\ref{Morse}) is the possibility of analytic progress.

For two particles interacting according to (\ref{Morse}) the ground-state 
energy is exactly known to be 
$E_{2} = -\epsilon [1 -  {\hbar/ 2 l (m\epsilon)^{1/2}}]^{2}$ 
\cite{Morse}.  This implies that in the many-body case one of the possible 
phases of the system is a {\it diatomic} gas (a collection of infinitely far
 separated dimers) with the reduced energy function 
\begin{equation}
\label{dimer}
E^{*}_{dimer}(\lambda_{0}) = - {1\over 2}(1 - {\pi \lambda_{0}\over \sqrt 2})^{2},
\end{equation} 
valid for $\lambda_{0} \le \lambda_{02} = \sqrt 2/\pi$;  the factor of $1/2$ 
accounts for two particles in the dimer.  As $\lambda_{0}$ 
approaches $\lambda_{02}$ from below, the dimer size diverges, and at 
$\lambda_{0} = \lambda_{02}$ a second-order dissociation transition into the 
monoatomic gas discussed earlier takes place.  The asymmetry of the 
interaction potential is responsible for the disappearance of the two-body 
bound state for sufficiently strong zero-point motion.         

The diatomic gas might be the ground-state of the system for intermediate
$\lambda_{0}$ but for sufficiently small $\lambda_{0}$ a condensed phase must 
have the lowest energy.  For molecular systems in general 
\cite{Ashcroft&Mermin} pair interactions decay rapidly with interparticle 
separation.  As a result, the physics of the condensed phase is dominated by
nearest-neighbor interactions.  Therefore in what follows in describing the
one-dimensional condensed phase we restrict ourselves to nearest-neighbor 
interactions.  Corrections coming from ignoring distant neighbors will be
marginally small provided the interaction range $l$ is substantially 
smaller than average interparticle spacing (bond length).

In the classical limit, $\lambda_{0} = 0$, the ground-state of the system is 
a one-dimensional crystal of particles with lattice spacing $H_{0}$ given by 
the minimum of the two-body Morse potential (\ref{Morse}).  Indeed, the 
energy per particle for the crystal, $-\epsilon$, is twice as negative as 
that for the diatomic gas.  

A condensed phase is also the ground-state of the system for a not very 
large $\lambda_{0}$.  The quantitative theory of the properties of this 
phase is developed below.     

\section{Luttinger liquid phase and its properties}

When zero-point motion is present, the asymmetry of the pair interaction 
potential about its minimum causes quantum expansion.  As a result the 
average bond length $H$ in the condensed state exceeds 
its classical, $\lambda_{0} = 0$, counterpart $H_{0}$.  The low-energy 
dynamics of the system is described by the harmonic action \cite{Luttinger}
\begin{equation}
\label{haction}
S_{harm} = {\rho \over 2} \int dx d\tau [({\partial u \over \partial \tau})^{2} + 
c^{2}({\partial u \over \partial x})^{2}],
\end{equation}
where $u(x,\tau)$ is the particle displacement field, $\rho = m/H$ is the 
mass density, and $c$ is the sound velocity 
\cite{Ashcroft&Mermin}
\begin{equation}
\label{sound}
c^{2} = {H^{2}\over m} {\partial ^{2} E (h = H)\over \partial h^{2}},
\end{equation}   
where $E(h)$ is the ground-state energy per particle as a function of 
(one-dimensional) volume  per particle $h$, and the derivative is evaluated at 
the equilibrium interparticle spacing $H$.  The function $E(h)$ can be also 
viewed as an effective pair interaction renormalized by zero-point motion away 
from its classical form (\ref{Morse}).  

The Feynman path integral formulation of quantum mechanics \cite{Kogut} allows 
us to view the action (\ref{haction}) as a Hamiltonian for a classical 
two-dimensional crystal of line objects (world lines of underlying particles) 
running in the imaginary time direction.  In this correspondence zero-point 
motion plays a role of thermal fluctuations.  But this is exactly 
the context of applicability of the Mermin-Wagner-Hohenberg theorem 
\cite{lro}: if $n(x,\tau)$ is the instantaneous number density, then 
long-wavelength low-energy quantum fluctuations captured by the action
(\ref{haction}) destroy long-range positional order of the particles.  The 
only allowed many-body  bound state is a uniform density 
phase, $<n(x,\tau)> = H^{-1}$ ($<>$ stands for the expectation value), with 
algebraic decay of density correlations \cite{Luttinger}:
\begin{equation}
\label{algebra}
<n(x,\tau)n(0,0)> - H^{-2} \propto {{\cos}(2\pi x/H)\over (x^{2} 
+ c^{2}\tau^{2})^{g}},  
\end{equation}
where the exponent $g$ is given by
\begin{equation}
\label{Lutg}
g = {\pi \hbar \over \rho c H^{2}}
\end{equation} 
The large distance/time behavior of the density-density correlation function 
(\ref{algebra}) is the hallmark of the Luttinger liquid.  In order to 
compute the range of existence of the Luttinger liquid phase, its bond length 
$H$, sound velocity $c$, and correlation exponent $g$ we need to go beyond 
harmonic approximation.

Our calculation relies on the assumption underlying the harmonic description
(\ref{haction}) (and to be verified later) that as long 
as the Luttinger liquid is stable, the ratio of the typical fluctuation of the 
bond length to the bond length itself remains small despite individually both 
these quantities are increasing functions of $\lambda_{0}$ (\ref{lambda_{0}}). 
Then every bond of the system can be viewed as a quantum-mechanical degree of 
freedom subject to the external potential $V(h)$ and placed in contact with
a bath of {\it harmonic} oscillators (\ref{haction}) corresponding to the 
rest of the system.  The single bond dynamics is thus given by the action:
\begin{equation}
\label{baction}
S_{bond} = {\rho \over 2} \int \limits_{bath} dx d\tau [({\partial u \over \partial \tau})^{2} + 
c^{2}({\partial u \over \partial x})^{2}] + \int d\tau V(h)
\end{equation}
The first integral is over all positions and times except for a small region 
near $x = 0$ where the bond in question is located.  This separates the system 
into two pieces, so that at all times the displacement field $u$ is 
discontinuous:  $u(x = +0, \tau) - u(x=-0,\tau) = h(\tau) - H_{0}$. The 
coupling between the segments of the system joined at the bond is given by 
the full pair potential $V(h)$, i. e. it goes beyond harmonic approximation.

The action (\ref{baction}) has been previously introduced in Ref.\cite{KS} to
describe tunneling-assisted fracture of a stretched one-dimensional chain.  
More generally this type of action corresponds to the Caldeira-Leggett model 
of coupling between a quantum-mechanical degree of freedom and an environment 
modeled by a reservoir of harmonic oscillators \cite{Caldeira&Leggett}.  
Since the last interaction term in (\ref{baction}) is restricted to a single 
spatial point, the bath degrees of freedom can be integrated out away from 
the bond with the result \cite{KS}:              
\begin{equation}
\label{waction}
S = {\rho c \over 4} \int \limits_{-\omega_{D}}^{\omega_{D}} 
{d\omega \over 2 \pi} |\omega| |h(\omega)|^{2} + \int d\tau V(h),
\end{equation}
where the subscript is dropped for brevity and the Fourier transform of the 
bond length field $h(\omega) = \int h(\tau) \exp(-\omega\tau)d\tau$ has been 
introduced \cite{note1}.  
The frequency cutoff $\omega_{D}$ setting the limits of integration in the 
first kinetic term of (\ref{waction}) is the one-dimensional Debye frequency 
$\omega_{D} = \pi c/H$ - the vibrational spectrum of the system is 
approximated  by the Debye model.    

The unusual $|\omega|$ dependence of the kinetic term of the action 
(\ref{waction}) is due to the many-body nature of the bond dynamics, and can be
understood heuristically by noticing that if the bond length oscillates with 
frequency $\omega$, then during one oscillation period $2\pi/|\omega|$ this 
disturbance propagates in both directions away from the bond a distance 
of order $c/|\omega|$.  Therefore the standard kinetic energy density, 
proportional to $\rho \omega^{2}$ should be multiplied by the size of the 
region $c/|\omega|$ affected by the motion thus reproducing the 
$\rho c |\omega|$ term of (\ref{waction}). 

The action (\ref{waction}) allows us in principle to compute how the bath 
degrees of freedom renormalize the properties of a given bond. The nonanalytic 
$|\omega|$ dependence in (\ref{waction}) guarantees that the
bond {\it cannot} renormalize the bath oscillators (whose properties are 
accumulated in the $\rho c$ combination).  This observation combined with the 
fact that all the bounds of the Luttinger liquid are equivalent provides us 
with a prescription on how to use (\ref{waction}) to solve the problem we are 
interested in:

The parameters $\rho$ and $c$ of (\ref{waction}) should be selected as 
initially unknown but fully renormalized properties of the Luttinger liquid.  
The reservoir degrees of freedom will renormalize the microscopic pair 
interaction $V(h)$ into a form which we will require to be identical (in the 
harmonic limit) to the rest of the chain.  This will determine the parameters 
of the Luttinger liquid and guarantee that the treatment is insensitive 
to the choice of the bond.       

Since the action (\ref{waction}) describes the dynamics of an arbitrary single 
bond of the system, it can be directly used to compute the energy
{\it per particle} of the original many-body problem.
   
\section{Zero-pressure analysis}  

For a quantitative analysis we use Feynman's \cite{Feynman} variational 
principle for the ground-state energy:
\begin{equation}
\label{var}
E \le E_{1} = E_{0} + (T/\hbar)<S - S_{0}>_{0}
\end{equation}
where $T$ is the temperature, and $\hbar/T$ has a meaning of the system size
in the $\tau$ direction; the $T = 0$ limit will be taken at the end.  The 
notation $<>_{0}$ denotes an expectation value computed using an arbitrary 
reference action $S_{0}$, and $E_{0}$ is the ground-state energy corresponding 
to $S_{0}$.  

This method has been remarkably successful in analyzing the roughening phase 
transition \cite{Saito}, multilayer adsorption phenomena \cite{Weeks}, 
wetting transitions \cite{LKZ}, the problem of quantum Brownian motion in a
periodic potential \cite{Fisher&Zwerger}, the Coulomb blockade problem 
\cite{KKQ}, and a variety of problems of quantum mechanics and quantum field 
theory \cite{Kleinert}.  

It is physically reasonable to select the trial action $S_{0}$ in a 
Gaussian form similar to that in \cite{Weeks}, \cite{LKZ}, and \cite{KKQ}:
\begin{equation}
\label{trialaction}
S_{0} = {\rho c \over 4} \int \limits_{-\omega_{D}}^{\omega_{D}} 
{d\omega \over 2 \pi} |\omega| |h(\omega)|^{2} + {K \over 2} \int d\tau (h - H)^{2}],
\end{equation}
where two variational parameters which include the familiar bond length $H$ 
and a new parameter $K$ (controlling the extent of fluctuations about $H$) are 
selected to minimize $E_{1}$ in (\ref{var}).  The stiffness parameter $K$ has 
a meaning of the curvature of the effective pair potential evaluated at its 
minimum $H$, and appearing in (\ref{sound}), 
$K = \partial^{2}_{h} E(h = H)$.   

Introducing $f = h - H$, the deviation of the bond length away from its 
equilibrium value $H$, the reduced root-mean-square (rms) fluctuation, 
$f^{*} = (<f^{2}>_{0}/l^{2})^{1/2}$, can be computed with the help of 
(\ref{trialaction}) as 
\begin{equation}
\label{msf}
f^{*} = (2 \lambda)^{1/2} \ln^{1/2}(1 + \gamma^{-1}),
\end{equation}
where
\begin{equation}
\label{lambda}
\lambda = {\hbar \over \pi \rho c l^{2}}
\end{equation}
is the dimensionless parameter quantifying the strength of zero-point motion
in the Luttinger liquid, and 
\begin{equation}
\label{gamma}
\gamma = {2K\over \omega_{D} \rho c}
\end{equation}
is the dimensionless counterpart of $K$. For the Morse pair interaction 
(\ref{Morse}) the classical sound velocity is 
$c_{0} = (H_{0}/l)(2\epsilon/m)^{1/2}$. It is then straightforward to verify 
that in the classical limit, $\hbar \rightarrow 0$, the quantum 
parameter $\lambda$ (\ref{lambda}) reduces to De Boer's  
number $\lambda_{0}$ (\ref{lambda_{0}}). 

In the Luttinger liquid the strength of zero-point motion is characterized by
the correlation exponent $g$ (\ref{Lutg}) which is related to 
$\lambda$ (\ref{lambda}) by
\begin{equation}
\label{goflambda}
g = \lambda \pi^{2}/Q^{2},
\end{equation}   
where $Q = H/l$ is the reduced bond length.  This relationship demonstrates that if the interaction range 
$l$ is known, then measuring the density-density correlation 
function (\ref{algebra}) will allow us to compute $\lambda$, and thus the 
rms fluctuation of the bond length (\ref{msf}). 
 
Using (\ref{trialaction}), (\ref{msf}), and (\ref{Morse}) the reduced upper 
bound $E^{*} = E_{1}/\epsilon$ entering (\ref{var}) can be computed as
\begin{eqnarray}
\label{varenergy1}
E^{*}(\gamma, Q)& =& v^{-1} \ln(1 + \gamma)-2e^{Q_{0} - Q}(1 + \gamma^{-1})^{\lambda}
\nonumber\\ 
& + & e^{2(Q_{0} - Q)}(1 + \gamma^{-1})^{4\lambda}, 
\end{eqnarray}
where 
\begin{equation}
\label{v}
v = {2 \pi \epsilon \over \hbar \omega_{D}}
\end{equation}
is related to the reduced Debye temperature as  
$\theta^{*} = \hbar \omega_{D}/\epsilon = 2\pi/v$.

The expression for $E^{*}$ should be minimized with respect to $\gamma$ and 
$Q$, and in case of multiple solutions the one minimizing (\ref{varenergy1}) 
must be selected.
 
Minimizing $E^{*}$ with respect to $Q$ we arrive at the expression for the 
reduced bond length which accounts for quantum expansion 
\begin{equation}
\label{qexpansion}
Q = Q_{0} + 3 \lambda \ln(1 + \gamma^{-1})
\end{equation}
Substituting this back into (\ref{varenergy1}) the expression for the reduced
energy $E^{*}$ can be written as   
\begin{equation}
\label{varenergy2}
E^{*}(\gamma) = v^{-1}\ln(1 + \gamma) 
- (1 + \gamma^{-1})^{-2\lambda},
\end{equation}

Minimizing (\ref{varenergy2}) with respect to $\gamma$ we find 
\begin{equation}
\label{gammaeq1}
\gamma = 2v \lambda (1 + \gamma^{-1})^{-2\lambda}
\end{equation}

\subsection{Approximate solution: $\lambda = \lambda_{0}$}

First we look at a simplified version of the original problem when only one 
bond of the system is subject to the Morse potential (\ref{Morse}) while 
the rest of the chain is harmonic.  This situation is described by the action
\begin{eqnarray}
\label{wetting}
S &=& {\rho_{0} c_{0} \over 4} \int \limits_{-\omega_{D0}}^{\omega_{D0}} 
{d\omega \over 2 \pi} |\omega| |h(\omega)|^{2}\nonumber\\ 
&+& \int d\tau(-A e^{-h/l} + B e^{-2h/l}),
\end{eqnarray}  
where $\rho_{0} = m/H_{0}$, $c_{0} = (H_{0}/l)(2\epsilon/m)^{1/2}$, and 
$\omega_{D0} =\pi c_{0}/H_{0}$ assume their classical values.  The 
parameters $A$ and $B$ are selected so that in the harmonic approximation the 
Morse bond is  identical to the rest of the chain.  

If the imaginary time coordinate $\tau$ is viewed as a fictitious space 
variable then (\ref{wetting}) can be recognized as an effective Hamiltonian \
defining a classical statistical mechanics problem \cite{Kogut}.  If the bond 
field $h$ is identified with an interface height and zero-point motion with 
thermal fluctuations, then this problem is a one-dimensional analog of the 
critical wetting problem \cite{LKZ}, \cite{BHL}.  

For finite $\lambda_{0}$ zero-point motion softens and lengthens the Morse 
bond while leaving the harmonic part of the chain intact. This is described by 
Eqs.(\ref{qexpansion})-(\ref{gammaeq1}) with $\lambda = \lambda_{0}$ and 
$v = v_{0} = 1/(\pi \lambda_{0})$.  Specifically, Eqs.(\ref{varenergy2}) and 
(\ref{gammaeq1}) turn into:
\begin{equation}
\label{varenergy3}
E^{*}(\gamma) = \pi [\lambda_{0} \ln(1 + \gamma) - {\gamma \over 2}]
\end{equation}
\begin{equation}
\label{gammaeq2}
\gamma ={2 \over \pi}(1 + \gamma^{-1})^{-2\lambda_{0}}
\end{equation}
First we note that $\gamma = 0$ is always a solution to (\ref{gammaeq2}) with 
$E^{*} = 0$ which describes two segments of the chain infinitely far apart
from each other.  As $\lambda_{0}$ 
increases away from the classical value $\lambda_{0} = 0$, the parameter 
$\gamma$ in (\ref{gammaeq2}) monotonically decreases from $\gamma = 2/\pi$ 
vanishing at $\lambda_{0} = 1/2$.  For $\lambda_{0} > 1/2$ only $\gamma = 0$ 
solves Eq.(\ref{gammaeq2}).  When $\lambda_{0}$ approaches the critical value 
of $1/2$ from below we have 
\begin{equation}
\label{gammacrit}
\gamma \simeq e^{-{\ln(\pi/2)\over 1 - 2\lambda_{0}}}
\end{equation}      
Correspondingly, the reduced energy (\ref{varenergy3}) monotonically increases
with $\lambda_{0}$: in the classical limit, $\lambda_{0} \rightarrow 0$, it 
rises linearly with $\lambda_{0}$ according to $E^{*} = - 1 
+ \lambda_{0}[\pi \ln(1 + 2/\pi) + 2\ln(1 + \pi/2)] \simeq -1 + 
3.4361\lambda_{0}$.  Since $\lambda_{0} \sim 1/m^{1/2}$, 
Eq. (\ref{lambda_{0}}), the energy $E^{*}$ is {\it not} analytic as 
$1/m \rightarrow 0$, which is the relevant quantity \cite{Anderson&Palmer}.  
This is expected because the crystal ($\lambda_{0} = 0$) is qualitatively 
different from the Luttinger liquid (finite $\lambda_{0}$).  Upon approaching 
$\lambda_{0} =1/2$ from below the reduced energy vanishes as
\begin{equation}
\label{energycrit}
E^{*} \simeq - {\pi \over 2}(1 - 2\lambda_{0}) e^{-{\ln(\pi/2)\over 1 - 2\lambda_{0}}}    
\end{equation}

As $\lambda_{0}$ approaches $1/2$ the bond length diverges and for 
$\lambda_{0} \ge 1/2$ the two segments of the chain are infinitely far apart 
from each other. 

The behavior of the reduced bond length and its rms fluctuation just below 
the phase transition can be found by combining (\ref{msf}) and 
(\ref{qexpansion}) with (\ref{gammacrit}):
\begin{equation}
\label{divergence}
Q \simeq  {3 \ln(\pi/2) \over 2 (1 - 2 \lambda_{0})},~~~~f^{*} \simeq 
{ \ln^{1/2}(\pi/2) \over (1 - 2 \lambda_{0})^{1/2}}
\end{equation}
We note that although both of these quantities diverge upon approaching the 
phase transition, the {\it relative} fluctuation, $f^{*}/Q$, vanishes. Thus 
fluctuating segments of the chain never overlap and our treatment is 
consistent.  

The essential singularities (\ref{gammacrit}) and (\ref{energycrit}) 
at $\lambda_{0} = 1/2$ as well as the divergences (\ref{divergence}) parallel 
those found in the context of wetting transitions \cite{LKZ}.

The most valuable feature of the variational approach is its nonperturbative 
nature.  The accuracy of variational predictions depends on how close is the 
variational guess to the physical reality.  For the problem defined by the
action (\ref{wetting}) the accuracy of our approach can be assessed and the 
special role played by $\lambda_{0} = 1/2$ can be re-established by using a 
renormalization-group method. 

\subsection{Perturbative renormalization-group treatment: 
$\lambda = \lambda_{0}$}

Following the argument of Br{\'e}zin, Halperin, and Leibler originally 
given in the classical context of wetting transitions \cite{BHL} we treat the 
Morse potential term in ({\ref{wetting}) as a perturbation.  Then the 
lowest-order renormalization-group equations have the form:
\begin{equation}
\label{rgeqab}
{d\ln A \over d\ln(\omega_{D0} \zeta)} = (\lambda_{0} + 1), ~~~  
{d\ln B \over d\ln(\omega_{D0} \zeta)} = (4\lambda_{0} + 1),  
\end{equation}
where $\zeta$ is the running scale in the $\tau$ direction, and the equations 
describe how the Morse parameters $A$ and $B$ renormalize upon (i) successive 
integration out of high-frequency modes (first terms) followed by 
(ii) scaling transformation which restores the cutoff to its original value 
(second terms).  Instead of following separate evolution of the coefficients 
$A$ and $B$, it is more appropriate to look at the depth of the Morse 
well (\ref{Morse}) $\epsilon = A^{2}/4B$ which is also proportional to 
the potential curvature at its minimum.  For its dimensionless counterpart $v$
(\ref{v}) Eqs.(\ref{rgeqab}) imply
\begin{equation}
\label{rgeqv}
{dv \over d\ln(\omega_{D0} \zeta)} = (1 - 2 \lambda_{0})v
\end{equation}   

The flow diagram corresponding to (\ref{rgeqv}) is sketched in Fig.2 where we 
also show the locus of initial conditions $v_{0} = 1/\pi \lambda_{0}$ of the 
model (\ref{wetting}).
\begin{figure}[htbp]
\epsfxsize=4.5in
\vspace*{-0.3cm}
\hspace*{-1.9cm}
\epsfbox{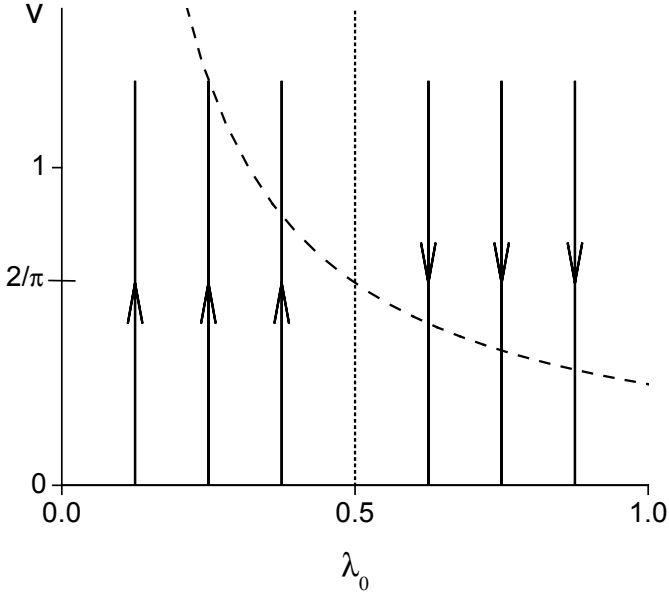}
\vspace*{0.1cm}
\caption{The flow diagram of the model (\ref{wetting}).  The arrows indicate 
the direction of the flow and the dashed line $v_{0} = 1/\pi \lambda_{0}$ is 
the locus of initial conditions.  The stable part of the $v = 0$, 
$\lambda_{0} > 1/2$ fixed line corresponds to two segments of the chain 
infinitely far away from each other.  For $\lambda_{0} < 1/2$ the parameter 
$v$ is relevant and the Morse bond joins the segments together.}  
\end{figure}

First we note that verticality of the flow lines is a rigorous property of the 
model (\ref{wetting}) as the $\rho_{0}c_{0}$ combination (and thus 
$\lambda_{0}$) does not renormalize to any order in 
$v$.  There are clearly two regimes separated by a phase transition at the 
critical value $\lambda_{0} = 1/2$.  For $\lambda_{0} > 1/2$ the parameter $v$
flows to zero which means that the chain is broken into two infinitely 
separated segments.  On the other hand, for $\lambda_{0} < 1/2$ the parameter 
$v$ grows under renormalization eventually leaving the perturbative regime 
$v \ll 1$.  This means that the Morse bond holds both segments of the chain 
together.  Upon approaching the critical value $\lambda_{0} = 1/2$ from below 
there is a divergent time scale $\xi$ (analog of a correlation length in 
standard critical phenomena) which can be found from the condition 
$v(\xi) \simeq 1$:
\begin{equation}
\label{xi}
\xi \simeq \omega_{D0}^{-1} e^{{\ln(1/v_{0})\over 1 - 2\lambda_{0}}}
\end{equation}
The one-dimensional nature of the problem then implies that the reduced 
curvature $\gamma$ (\ref{gamma}) of the renormalized potential vanishes 
as $\xi^{-1}$.  Similarly the critical behavior of the reduced bond 
length $Q$ can be found as 
$Q \simeq \ln[B(\xi)/A(\xi)] \simeq (3/2)\ln(\omega_{D0}\xi)$, while its rms
fluctuation is $f^{*} \simeq \ln^{1/2}(\omega_{D0}\xi)$.  With $\xi$ given by 
(\ref{xi}) and logarithmic accuracy these results 
coincide with their variational counterparts, Eq.(\ref{gammacrit}) and 
(\ref{divergence}).  If we set the bare parameter $v_{0}$ in (\ref{xi}) at 
$2/\pi$ (the crossing of the locus of initial conditions 
$v_{0} = 1/\pi \lambda_{0}$ and the $\lambda_{0} = 1/2$ line) the 
renormalization-group results would become identical to those of the 
variational approach.

From perturbative renormalization-group treatment alone we would not be able 
to make reliable statements about the phase transition at $\lambda_{0} =1/2$ as
the analysis is valid for $v_{0} \ll 1$ while for the problem in question 
(\ref{wetting}) one has $v_{0} = 2/\pi \simeq 0.64 < 1$ - it is on the border 
of applicability of perturbative theory.  However combining the above results
with the nonperturbatve variational analysis makes a strong case.   Since the 
latter produces the same answers in the region of parameters where 
renormalization-group results are less certain, we argue that the
variational solution of (\ref{wetting}) is very accurate in the range
of $\lambda_{0}$ between the classical limit $\lambda_{0} = 0$ and the 
dissociation transition $\lambda_{0} =  1/2$ which is described exactly.

As a side observation we note that the problem (\ref{wetting}) is most likely 
to be related to that of quantum Brownian motion of a particle in a {\it 
periodic} potential \cite{Fisher&Zwerger}.  Superficially the only similarity 
between the two is that (\ref{wetting}) can be also viewed as describing 
quantum Brownian motion in the Morse potential which has no periodicity.  

The similarity between the problems becomes noticeable if one inspects their 
treatments.  A comparison shows that our expression for 
variational energy (\ref{varenergy2}) is identical to that of the 
periodic version of the problem with the amplitude of the periodic potential 
proportional to our parameter $v$ (\ref{v}) and our $\lambda = \lambda_{0}$ 
corresponding to $1/2\alpha$ of Fisher and Zwerger \cite{Fisher&Zwerger}.  
With this identification renormalization-group equation (\ref{rgeqv}) 
coincides with its periodic counterpart \cite{Fisher&Zwerger}.  Both 
problems have delocalization transitions of the same universality class driven 
by zero-point motion.

We remind the reader that the action (\ref{wetting}) is an approximation to 
the original problem of the ground-state properties of the Luttinger liquid - 
only one bond is subject to the Morse potential while the rest of the system
is purely harmonic.  This chain is obviously {\it stiffer} than the original 
system - a {\it smaller} level of zero-point motion will be necessary to cause 
dissociation of the Luttinger liquid.  Therefore the results derived for the
model (\ref{wetting}) imply that {\it zero-pressure Luttinger liquid phase 
cannot exist for $\lambda_{0} \ge 1/2$}.   

We also note that the critical value $\lambda_{0} = 1/2$ is {\it larger} than 
the dimer dissociation threshold $\lambda_{02} = \sqrt 2/\pi \simeq 0.4502$.  
This can be understood qualitatively by noticing that the dynamics of the 
Morse bond joining two half-infinite harmonic segments is more inertial 
(and thus more classical) than that of the Morse dimer.  Therefore with the 
same underlying particles a {\it weaker} level of quantum fluctuations 
(i. e. {\it smaller} $\lambda_{0}$) suffices to break the dimer.

\subsection{Accurate solution} 

In order to compute the properties of the Luttinger liquid more accurately we 
have to impose the condition that all the bonds of the chain are equivalent.  
Then using definitions of De Boer's number $\lambda_{0}$ (\ref{lambda_{0}}) 
and its Luttinger liquid counterpart $\lambda$ (\ref{lambda}) the parameter 
$v$ (\ref{v}) can be calculated as $v = \lambda/\pi \lambda_{0}^{2}$; for 
$\lambda = \lambda_{0}$ it reduces to $v =1/\pi \lambda_{0}$ previously used in
approximate treatment of the problem.  Similarly the reduced curvature of the
effective pair potential (\ref{gamma}) can be computed with the help of 
Eq.(\ref{sound}) with the conclusion that $\gamma = 2/\pi$.  This is exactly 
what was previously found in the approximate analysis in the classical limit,
$\lambda_{0} =0$, when indeed all the bonds of the chain are equivalent.

Substituting $v = \lambda/\pi \lambda_{0}^{2}$, and $\gamma = 2/\pi$ in 
Eq.(\ref{varenergy1}) we arrive at the expression for the reduced energy as a 
function of dimensionless ``volume'' per particle $Q$:
\begin{eqnarray}
\label{eofqfinal}
E^{*}(Q)& =& (\pi \lambda_{0}^{2}/\lambda)  \ln(1 + 2/\pi)-2e^{Q_{0} - Q 
+\lambda \ln(1 + \pi/2)}
\nonumber\\ 
& + & e^{2(Q_{0} - Q) + 4\lambda\ln(1 + \pi/2)} 
\end{eqnarray}
Similarly Eqs.(\ref{msf}), (\ref{qexpansion})-(\ref{gammaeq1}) transform into
\begin{equation}
\label{rmsfinal}
f^{*} = (2 \lambda)^{1/2} \ln^{1/2}(1 + \pi /2)         
\end{equation}
\begin{equation}
\label{qexpansionfinal}
Q = Q_{0} + 3 \lambda \ln(1 + \pi/2)
\end{equation}
\begin{equation}
\label{varenergy4}
E^{*}_{LL} = e^{-2\lambda \ln(1 + \pi/2)} [\pi \lambda \ln(1 + 2/\pi) -
1]
\end{equation}
\begin{equation}
\label{gammaeq3}
\lambda_{0} = \lambda e^{-\lambda \ln(1 + \pi/2)}
\end{equation}
Eqs.(\ref{eofqfinal})-(\ref{varenergy4}) are the main results of this 
Section.   
We note that in view of Eqs.(\ref{algebra}) and (\ref{goflambda}) 
Eqs.(\ref{rmsfinal})-(\ref{varenergy4}) give the dependence of the reduced 
rms bond length fluctuation $f^{*}$, bond length $Q$, and
energy per particle of the Luttinger liquid $E^{*}_{LL}$ on the
experimentally measurable quantum parameter $\lambda$.

The dependence of De Boer's number $\lambda_{0}$ on $\lambda$ is given by 
Eq.(\ref{gammaeq3}); we are interested in the inverse dependence, 
$\lambda(\lambda_{0})$.

The right-hand side of (\ref{gammaeq3}) has a maximum at 
$\lambda_{s} = \ln^{-1}(1 + \pi/2) \simeq 1.0591$ of magnitude 
$\lambda_{0s} = e^{-1} \ln^{-1}(1 + \pi/2) \simeq 0.3896$.  For 
$\lambda_{0} < \lambda_{0s}$ Eq.(\ref{gammaeq3}) has two roots for $\lambda$ 
but only the smaller one is physical.  For $\lambda_{0} = \lambda_{0s}$ 
these two roots coincide, and for $\lambda_{0} > \lambda_{0s}$ 
Eq.(\ref{gammaeq3}) has no solutions - Luttinger liquid phase is no longer 
stable.  We note that the Luttinger liquid cannot sustain the level of 
zero-point motion stronger than that corresponding to
$\lambda_{0s} \simeq 0.3896$ which is {\it smaller} than the  dimer 
dissociation threshold $\lambda_{02} = \sqrt 2/\pi \simeq 0.4502$. 
\begin{figure}[htbp]
\epsfxsize=4.1in
\vspace*{-1.0cm}
\hspace*{-1.0cm}
\epsfbox{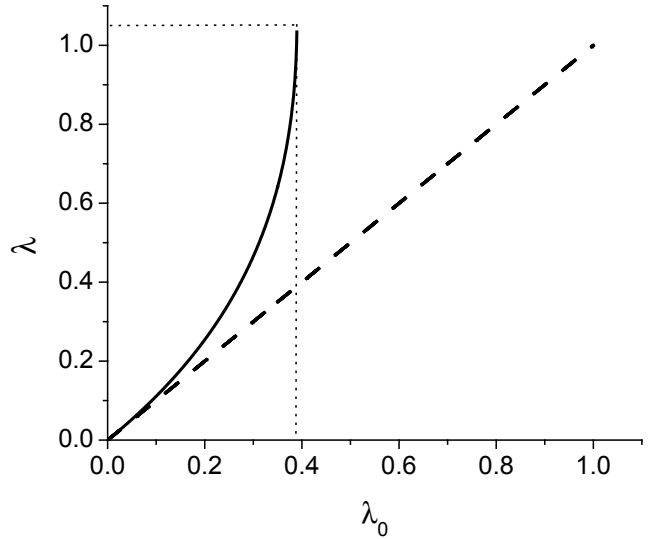}
\vspace*{0.1cm}
\caption{The dependence of the Luttinger liquid quantum parameter $\lambda$,
Eq.(\ref{lambda}) on De Boer's number $\lambda_{0}$, Eq.(\ref{lambda_{0}}).  
The dashed line is $\lambda = \lambda_{0}$.}     
\end{figure}
The $\lambda(\lambda_{0})$ dependence found by inverting Eq.(\ref{gammaeq3})
is shown in Fig.3 by a solid line. In the classical limit, 
$\lambda_{0} \rightarrow 0$, we have $\lambda \rightarrow \lambda_{0}$ as 
expected;  then the effective pair potential  (\ref{eofqfinal}) reduces to its
bare Morse form.  As $\lambda_{0}$ grows, the difference between 
$\lambda$ and $\lambda_{0}$ increases.  We note that De Boer's number 
$\lambda_{0}$ (\ref{lambda_{0}}) is always {\it smaller} than its Luttinger 
liquid counterpart $\lambda$ (\ref{lambda}) which reflects softening of the 
Luttinger liquid by zero-point motion.  The end point of the 
$\lambda(\lambda_{0})$ dependence which is the limit of stability of the 
liquid phase is a critical phenomenon - there the 
$\lambda(\lambda_{0})$ dependence has an infinite slope.  Upon approaching 
$\lambda_{0s}$ from below we find that $\lambda$
is given by 
\begin{equation}
\label{endpoint}
\lambda = \lambda_{s} 
- [{2e \over \ln(1 + \pi/2)} (\lambda_{0s} - \lambda_{0})]^{1/2}
\end{equation}  

Although both the reduced rms fluctuation $f^{*}$ (\ref{rmsfinal}) and bond 
length $Q$ (\ref{qexpansionfinal}) are increasing functions of the 
Luttinger liquid parameter $\lambda$ (\ref{lambda}), the relative 
fluctuation $f^{*}/Q$ has a maximum at $\lambda\ln(1 + \pi/2) = Q_{0}/3$.  
While the discussion of specific molecular substances will be postponed 
until Section VI, here we note that in our attempts to fit real two-body 
potentials into the Morse form the reduced classical bond length was 
always found to satisfy $Q_{0} \gtrsim 5$.  The relatively large value of 
$Q_{0}$ is a reflection of the strength of the short-distance overlap 
repulsion and weakness of the large-distance attraction in the pair 
interaction potential.  With $Q_{0} = 5$ (used hereafter for estimates) we 
find that $\lambda = (5/3)\ln^{-1}(1 + \pi/2) \simeq 1.7651$ at the maximum 
of the relative fluctuation.  But the Luttinger liquid cannot exist for 
$\lambda > \lambda_{s} = \ln^{-1}(1 + \pi/2) \simeq 1.0591$. 

Therefore the relative fluctuation, $f^{*}/Q$, reaches its maximal value at 
the border of existence of the Luttiner liquid, $\lambda_{s} \simeq 1.0591$, 
with the magnitude  not exceeding the level of 
about $0.18$.  This fact resembling Lindemann's empirical criterion of 
melting \cite{Lindemann} verifies that our description of the bond dynamics 
as due to coupling to a bath of {\it harmonic} oscillators is quantitatively 
correct.  The inequality  $f^{*}/Q \ll 1$ implies that underlying particles 
never come into close contact with each other - the effect of particle 
statistics is negligible.  In addition, having $Q_{0}$ substantially larger 
than unity justifies the nearest-neighbor interaction approximation in our 
treatment of the liquid.  
\begin{figure}[htbp]
\epsfxsize=4.1in
\vspace*{-1.0cm}
\hspace*{-1.8cm}
\epsfbox{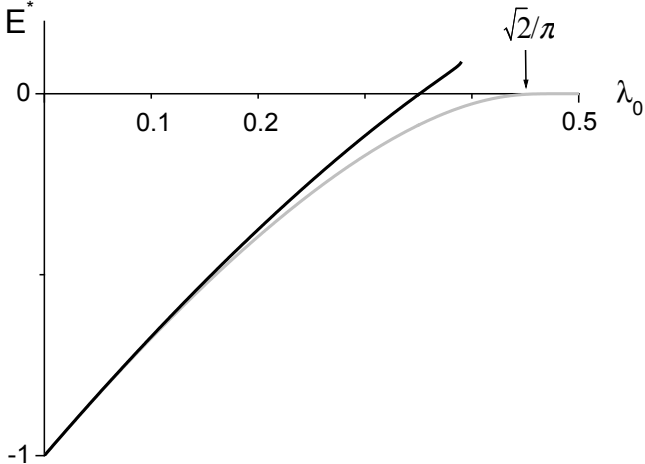}
\vspace*{-1.0cm}
\caption{The reduced energy per particle of the Luttinger liquid 
phase as a function of De Boer's number $\lambda_{0}$ (\ref{lambda_{0}}).  The 
lower grey scale curve corresponds to approximate solution described by 
Eqs.(\ref{varenergy3}) and (\ref{gammaeq2}) while the upper curve is an 
accurate solution given by Eqs.(\ref{varenergy4}) and (\ref{gammaeq3}). The 
arrow shows the location of the dimer dissociation threshold 
$\lambda_{02} = \sqrt 2/\pi$.}     
\end{figure}
The only remaining approximation which needs to be addressed is our replacement
of the vibrational spectrum of the system, $\omega(k) = (2c/H)|\sin(kH/2)|$, 
by the Debye model, $\omega(k) = c|k|$, valid for 
$\omega \le \omega_{D} = \pi c/H$.  Both spectra are fairly close to each 
other:  they coincide in the long-wavelength limit, $kH \ll 1$, and end at the
edges of the first Brillouin zone, $k = \pm \pi /H$.  Physically the chain 
with the Debye spectum is less susceptible to short-wavelength fluctuations 
than the original system - the Debye frequency $\omega_{D}$ is $\pi/2$ times
larger than the maximal allowed frequency.  This difference necessary to have
the correct number of degrees of freedom will only have a marginally small 
effect on final results because of the dominant role played in one dimension 
by low-energy long-wavelength fluctuations where the Debye approximation 
becomes exact. 
\begin{figure}[htbp]
\epsfxsize=3.7in
\vspace*{-0.5cm}
\hspace*{-0.8cm}
\epsfbox{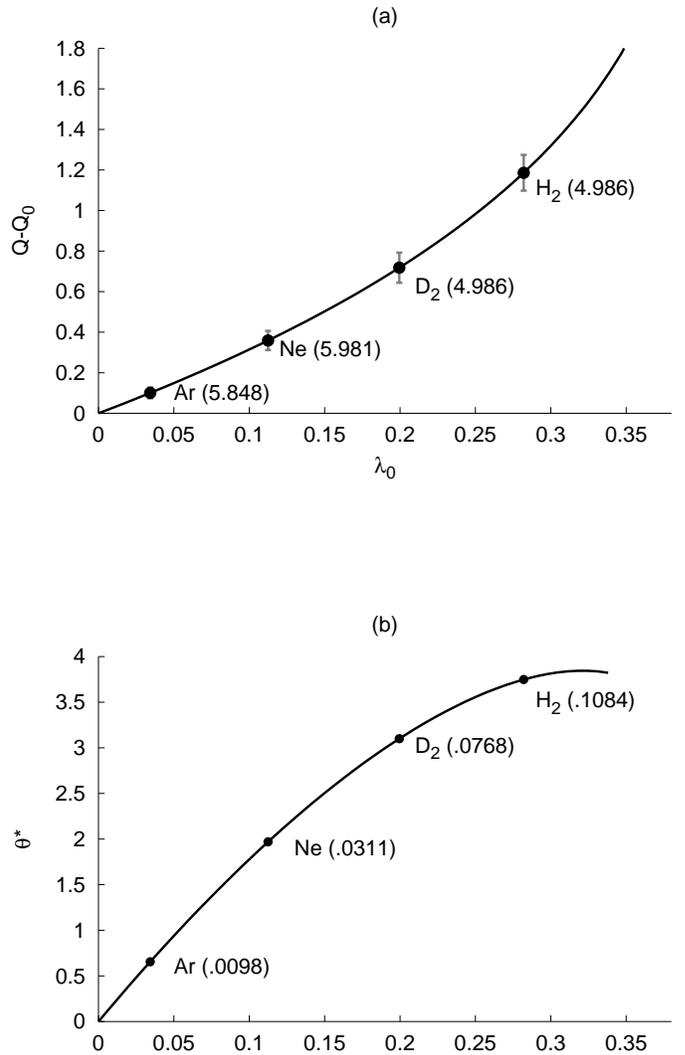}
\vspace*{-0.5cm}
\caption{The reduced quantum expansion $Q - Q_{0}$ (a), and the Debye 
temperature $\theta^{*}$ (b) of the Luttinger liquid as
functions of De Boer's number $\lambda_{0}$ (\ref{lambda_{0}}). Various 
molecular substances are shown as solid dots on the curves.  The numbers in 
the brackets on the quantum expansion graph are classical reduced 
bond lengths $Q_{0}$ while the size of vertical bars equals the relative bond 
fluctuation.  The numbers in the brackets on the Debye temperature graph are 
Luttinger liquid exponents $g$ (\ref{Lutg}).}     
\end{figure}
The upper curve of Fig.4 shows the reduced ground-state energy per particle of 
the Luttinger liquid as a function of De Boer's number 
$\lambda_{0}$ (\ref{lambda_{0}}) found by combining Eqs.(\ref{varenergy4}) 
and (\ref{gammaeq3}).  Qualitatively similar dependence was found in 
Ref.\cite{direct}; our energy is about $10$ percent lower.  For the 
purpose of comparison the lower grey scale curve of Fig.4 shows the 
approximate reduced energy per particle, Eqs.(\ref{varenergy3}) and 
(\ref{gammaeq2}), which was argued to constrain the 
$E^{*}_{LL}(\lambda_{0})$ dependence from below. The quantitative difference 
between the curves is not very large and becomes noticeable only for 
$\lambda_{0} > 0.15$.  Since variational analysis always constrains the 
ground-state energy from above, the true $E^{*}_{LL}(\lambda_{0})$ dependence 
must be sandwiched between the curves of Fig.4.  We expect however that 
$E^{*}_{LL}(\lambda_{0})$ is well-approximated by Eqs.(\ref{varenergy4}) and 
(\ref{gammaeq3}) (the upper curve of Fig.4).

Variation of other properties of the Luttinger liquid with De Boer's number
$\lambda_{0}$ (\ref{lambda_{0}}) can be readily calculated. The expression for 
the reduced sound velocity $c^{*}= c/c_{0}$ can be found to be 
\begin{eqnarray}
\label{reducedsound}
c^{*}& =& \lambda_{0}Q/\lambda Q_{0}
\nonumber\\
& =& e^{-\lambda \ln(1 + \pi/2)}[1 +  3 (\lambda/Q_{0}) \ln(1 + \pi/2)],
\end{eqnarray} 
which together with (\ref{gammaeq3}) parametrically determine 
$c^{*}(\lambda_{0})$. In the range of interest $Q_{0} \gtrsim 5$ the reduced sound
velocity is a monotonically decreasing function of $\lambda_{0}$.  At the 
boundary of existence of the Luttinger liquid the reduced sound velocity 
reaches its minimal (and finite) value and has a square-root singularity 
implied by Eq.(\ref{endpoint}). 

Similarly the dependence of the Luttinger liquid exponent $g$ (\ref{Lutg}) on
$\lambda_{0}$ (\ref{lambda_{0}}) can be found by combining 
Eqs.(\ref{goflambda}), (\ref{qexpansionfinal}) and (\ref{gammaeq3}).  Again for
$Q_{0} \gtrsim 5$ the correlation exponent $g$ is a monotonically increasing 
function of $\lambda_{0}$ reaching its maximal value at the boundary of 
existence of the liquid phase.

The $\lambda_{0}$ dependence of the reduced Debye temperature $\theta^{*}$ 
is determined by 
\begin{equation}
\label{Debye}
\theta^{*} = 2\pi/v = 2\pi^{2}\lambda e^{-2\lambda \ln(1 + \pi/2)},
\end{equation}
combined with Eq.(\ref{gammaeq3}).  The right-hand side of (\ref{Debye}) has a
maximum at $\lambda = \lambda_{s}/2 = 0.5 \ln^{-1}(1 + \pi/2) \simeq 0.5295$ 
which is inside the range of existence of the liquid phase.  In view of 
(\ref{gammaeq3}) this corresponds to De Boer's number 
$\lambda_{0} = 1/[2e^{1/2}\ln(1 + \pi/2)] \simeq 0.3212$ which is the location
of the maximum of the $\theta^{*}(\lambda_{0})$ dependence.

Fig.5 shows the dependence of the reduced quantum expansion 
$Q - Q_{0}$ and Debye temperature $\theta^{*}$ on De Boer's number 
$\lambda_{0}$.  The quantum expansion is  found by combination of 
Eqs.(\ref{qexpansionfinal}) and (\ref{gammaeq3}); it is merely a magnified 
$\lambda(\lambda_{0})$ curve of Fig.3.  The $\theta^{*}(\lambda_{0})$ 
dependence (Fig.5b) is constructed by combining Eqs.(\ref{Debye}) and 
(\ref{gammaeq3}).  Both properties are linear functions of De Boer's number 
in the $\lambda_{0} \rightarrow 0$ limit.

\subsection{Zero-pressure phase diagram}

In order to construct the phase diagram of the system we need to compare all 
the branches of the reduced energy function (\ref{qtcs}); for given De Boer's 
number $\lambda_{0}$ the branch with lowest  $E^{*}(\lambda_{0})$ singles out 
the ground-state.  The outcome is shown in Fig.6 where the reduced energy per 
particle of the Luttinger liquid, Eqs.(\ref{varenergy4}) and 
(\ref{gammaeq3}), is drawn together with those for diatomic, Eq.(\ref{dimer}), 
and monoatomic, $E^{*}_{mono}(\lambda_{0}) = 0$, gases.  

We see that as $\lambda_{0}$ increases away from the classical limit, 
$\lambda_{0} = 0$, the reduced energy per particle of the Luttinger liquid
increases, and at $\lambda_{0ev} \simeq 0.3365$ a crossing with the
dimer gas energy curve, Eq.(\ref{dimer}), takes place - liquid phase 
evaporates into diatomic gas via discontinuous transition.  We
note that the Luttinger liquid can still coexist (as a metastable state) with 
gas phases in the narrow range $0.3365 \lesssim \lambda_{0} \lesssim 0.3896$.  
This is shown by the grey scale part of the energy curve.
\begin{figure}[htbp]
\epsfxsize=3.55in
\vspace*{0.0cm}
\hspace*{-0.5cm}
\epsfbox{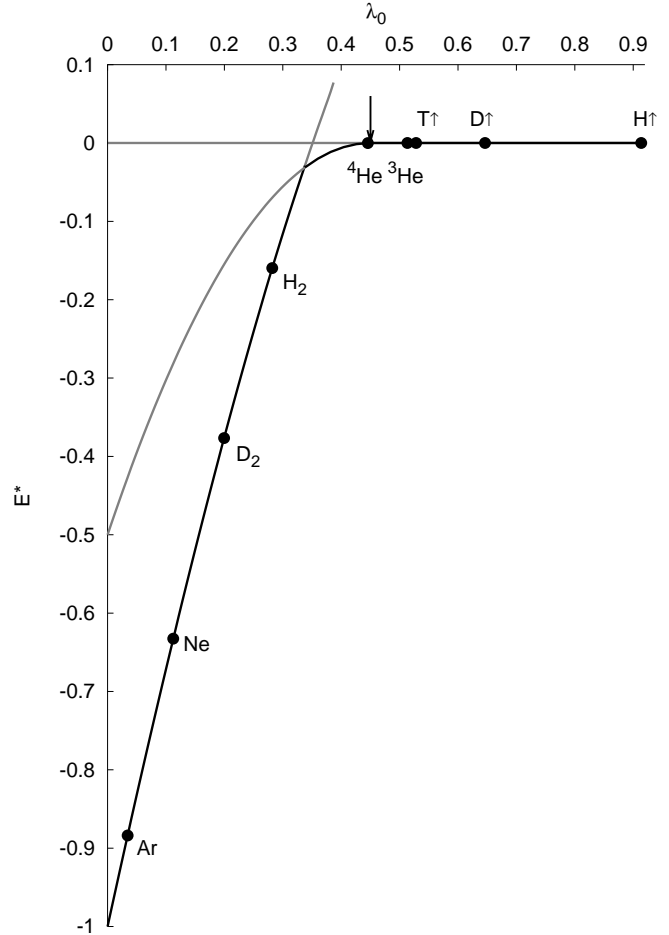}
\vspace*{0.5cm}
\caption{The dependencies of the reduced energy per particle on De Boer's 
number $\lambda_{0}$ (\ref{lambda_{0}}) for various phases. The bold 
parts of the curves correspond to ground states of the system while the grey
scale segments indicate metastable states.  The loci of a series of substances 
are shown by solid dots, and the arrow pointing down is the dimer dissociation 
threshold.}     
\end{figure}

The diatomic gas is the ground-state of the system in the range 
$0.3365 \lesssim \lambda_{0} < \sqrt 2/\pi \simeq 0.4502$:  at 
$\lambda_{02} = \sqrt 2/\pi$ it undergoes a continuous
dissociation transition into a monoatomic gas.  For 
$\lambda_{0} > \sqrt 2/\pi$ quantum fluctuations are too strong and no
bound state can exist:  a monoatomic gas is the only possible state of
the system.

In determining the ground-state one needs to take into account all possible 
competing phases of the system in question.  We however neglected the 
possibility that trimers (and generally $N$-atomic molecules) may come into
play.   This is because we only know how to treat the $N = 1, 2,
\infty$ cases in a controlled  fashion.   Although it appears unlikely, we 
cannot rule out that for sufficiently large $\lambda_{0}$, $N > 2$-atomic 
gases might become relevant;  resolving this issue is left for future study.

A critical reader still may argue that our prediction of the diatomic gas 
phase is an artifact of the variational treatment - the  exact energy curve 
corresponding to the Luttinger liquid may go lower then what Fig.6 shows.  If 
this is the case, then the dimer gas ground state may disappear altogether.  

A finite-pressure treatment described next provides additional evidence 
that physics is incomplete with only liquid and monoatomic gas present.
 
\section{Finite pressure}

At zero pressure there is a qualitative difference between a liquid which is a 
bound many-body state and a gas which is a collection of infinitely far 
separated particles.  Arbitrarily small confining pressure necessarily 
brings a gas to a finite density.  As a result the two gas phases previously 
discussed turn into Luttinger liquids.  The difference between the ``parent'' 
Luttinger liquid and what used to be a monoatomic gas becomes merely 
quantitative - they will have differing densities, sound velocities, 
correlation exponents and other properties.  The liquid of dimers is more 
complicated as in addition its oscillation spectrum will have an extra 
optical branch.
    
To avoid confusion we will keep referring to these pressure-induced Luttinger 
liquids as gases. As the pressure and De Boer's number 
$\lambda_{0}$ change, they may undergo gas-gas and gas-liquid transitions.  At 
sufficiently large pressure and $\lambda_{0}$ the quantitative difference 
between the gases and the liquid must disappear.   

In what follows we will not be able to discuss these transitions as at the 
moment it is unclear how to describe the effect of pressure on the diatomic 
gas phase in a controlled fashion.  On the other hand, generalization of our 
formalism to monoatomic gas and Luttinger liquid is straightforward.  Thus we 
will restrict ourselves to finding ranges of existence of these phases.

The difference between the liquid and gas becomes most extreme at 
{\it negative} pressure.  Here the liquid may still exist as a metastable 
state while the gas phase is impossible.
     
In one dimension the pressure has dimensionality of a force.  Assume our 
system is compressed by a constant force $p$ applied to its 
ends.  The system responds by exerting an {\it outward} force of magnitude $p$ 
on the compressing agent which corresponds to the definition of 
positive pressure.  Similarly, if the system is stretched by an external force,
it responds by exerting an {\it inward} force on the stretching agent which 
corresponds to the definition of negative pressure.  These two cases will be 
distinguished by the sign of $p$.  

Since every particle of the system is in mechanical equilibrium,  the whole 
effect of pressure translates into replacing the bond potential 
$V(h)$ by $V(h) + ph$, the total potential energy in the external field 
\cite{KS}.  Then the finite-pressure analog of Eq.(\ref{varenergy1}) becomes
\begin{eqnarray}
\label{pvarenergy1}
E^{*}(\gamma, Q)& =& v^{-1} \ln(1 + \gamma)-2e^{Q_{0} - Q}(1 + \gamma^{-1})^{\lambda}
\nonumber\\ 
& + & e^{2(Q_{0} - Q)}(1 + \gamma^{-1})^{4\lambda} + p^{*}Q, 
\end{eqnarray}  
where $p^{*} = pl/\epsilon$ is the reduced pressure. Eq.(\ref{pvarenergy1}) 
should be minimized with 
respect to $\gamma$ and $Q$, and then $v = \lambda/\pi\lambda_{0}^{2}$ and 
$\gamma = 2/\pi$ substituted in the outcome will guarantee translational 
invariance.  The results are finite pressure analogs of Eqs.(\ref{eofqfinal}),
(\ref{qexpansionfinal}), and (\ref{gammaeq3}):
\begin{eqnarray}
\label{peofqfinal}
E^{*}(Q)& =& (\pi \lambda_{0}^{2}/\lambda)  \ln(1 + 2/\pi)-2e^{Q_{0} - Q 
+\lambda \ln(1 + \pi/2)}
\nonumber\\ 
& + & e^{2(Q_{0} - Q) + 4\lambda\ln(1 + \pi/2)} + p^{*}Q,
\end{eqnarray}
\begin{eqnarray}
\label{pqexpansionfinal}
Q& =& Q_{0} + 3 \lambda \ln(1 + \pi/2)
\nonumber\\
& -&\ln[(1 + \sqrt{1 
+ 2p^{*}e^{2 \lambda \ln(1 + \pi/2)}})/2],
\end{eqnarray} 
\begin{eqnarray}
\label{pgammaeq3}
\lambda_{0}& =& \lambda e^{-\lambda\ln(1 + \pi/2)}
[(1+ 2p^{*}e^{2\lambda\ln(1 + \pi/2)}
\nonumber\\
& + & \sqrt{1 
+ 2p^{*}e^{2 \lambda \ln(1 + \pi/2)}})/2]^{1/2}
\end{eqnarray}
The expressions for the reduced rms fluctuation (\ref{rmsfinal}) and Debye
temperature (\ref{Debye}) as functions of $\lambda$ remain the same while the
$\lambda(\lambda_{0})$ dependence is determined by (\ref{pgammaeq3}).  
Similarly the reduced sound velocity is given by the first representation of 
Eq.(\ref{reducedsound}) with $Q$ and $\lambda$ determined by 
Eqs.(\ref{pqexpansionfinal}) and (\ref{pgammaeq3}), respectively. 

We note that the parameter $\lambda$ (\ref{lambda}) now accounts for both the 
effects of pressure and zero-point motion.

\subsection{Classical limit}

In the classical limit $\lambda_{0}, \lambda \rightarrow 0$ and $p^{*} > 0$ 
the 
position of the minimum of (\ref{peofqfinal}) given by 
(\ref{pqexpansionfinal}) naturally shifts to values smaller than $Q_{0}$.  
In addition Eq.(\ref{pqexpansionfinal}) predicts that at a very large pressure 
$p^{*}_{0} \simeq 2e^{2Q_{0}} = 2e^{10} \simeq 44000$ the bond length 
vanishes.  This conclusion is an artifact because for small interparticle 
separation the Morse potential underestimates the true strength of overlap 
repulsion - the bond length can only go to zero in the limit of infinite 
pressure.  This flaw implies that only $p^{*} \ll p^{*}_{0}$ results are 
credible which is not really restrictive as $p^{*}_{0}$ is unrealistically 
large. 

The pressure dependence of the reduced sound velocity follows from the first 
representation of (\ref{reducedsound}), and Eqs.(\ref{pqexpansionfinal}) and
(\ref{pgammaeq3}).         

For not very large negative pressure the pair potential (\ref{peofqfinal}) has
a minimum given by (\ref{pqexpansionfinal}) and a maximum - the ``broken'' 
ground state of the chain is separated from the metastable stretched crystal 
by a potential barrier.  As the magnitude of the pressure increases, the 
amplitude of the barrier decreases, and at the classical limit of mechanical 
stability, $p^{*}_{ms} = -1/2$, the barrier disappears altogether - no bound 
state can exist for $p^{*} < -1/2$.  As the pressure approaches the limit of 
mechanical stability from above, the reduced bond length tends to 
$Q = Q_{0} + \ln2$ while the reduced sound velocity vanishes as 
$(1 + 2p^{*})^{1/4}$ in agreement with general arguments of Ref.\cite{KS}.

\subsection{Quantum case}

With quantum effects included, the pressure at which the bond length vanishes
(and the Morse potential approximation fails) will be even larger than its 
classical counterpart as zero-point motion counteracts the compression.  
Therefore the restriction $p^{*} \ll p^{*}_{0}$ remains unchanged in the 
quantum case.

Since quantum fluctuations lead to the expansion and softening of the liquid 
phase, a smaller in magnitude negative pressure will suffice to destabilize 
the liquid - the dependence of the limit of mechanical stability on De Boer's 
number (\ref{lambda_{0}}), $p^{*}_{ms}(\lambda_{0})$, should be a monotonically
increasing function of $\lambda_{0}$ satisfying $p^{*}_{ms}(0) = - 1/2$ 
(classical limit) and $p^{*}_{ms}(\lambda_{0s}) = 0$ (zero-pressure limit of 
stability of the Luttinger liquid).
\begin{figure}[htbp]
\epsfxsize=4.1in
\vspace*{-1.0cm}
\hspace*{-1.0cm}
\epsfbox{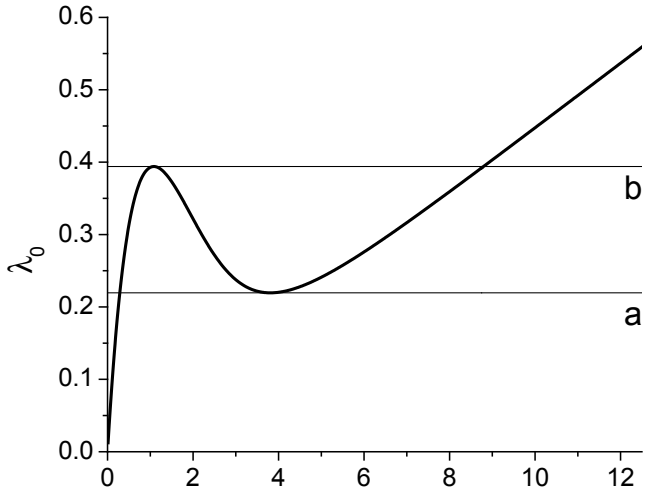}
\vspace*{-0.5cm}
\caption{The $\lambda_{0}(\lambda)$ dependence, Eq.(\ref{pgammaeq3}), for 
suffifciently small positive pressure ($p^{*} = 0.002$ is shown).  The lines 
of fixed $\lambda_{0} = \lambda_{0a,b}$ are the limits of stability of gas 
and liquid, respectively.}
\end{figure} 
For $p^{*}= 0$ the right-hand-side of (\ref{pgammaeq3}) vanishes both at 
$\lambda = 0$ and $\lambda = \infty$ reaching a maximum at 
$\lambda = \lambda_{s}$ which determined the limit of stability of the liquid 
in the zero-pressure case (see Section IVC).  For finite positive pressure and 
$\lambda \rightarrow \infty$ the right-hand-side of (\ref{pgammaeq3}) behaves 
as $p^{*1/2}\lambda$ which implies that for not very large $p^{*}$ the
$\lambda_{0}(\lambda)$ dependence is a nonmonotonic function which has both
a maximum and a minimum.  The position of the minimum shifts to infinity as
$p^{*} \rightarrow 0$.  An example of the $\lambda_{0}(\lambda)$ dependence, 
Eq.(\ref{pgammaeq3}), for sufficiently small positive pressure is displayed 
in Fig.7 where we also show two lines of constant $\lambda_{0}$ to help 
identify possible phases of the system. 

For fixed small positive pressure and $\lambda_{0} < \lambda_{0a}$ 
Eq.(\ref{pgammaeq3}) has a unique solution for $\lambda$ describing the 
liquid.  For $\lambda_{0a} < \lambda_{0} < \lambda_{0b}$ Eq.(\ref{pgammaeq3}) 
has three solutions.  Out of them only the smallest (corresponding to liquid) 
and the largest (corresponding to gas) are physical.  For 
$\lambda_{0} > \lambda_{0b}$ there is only one solution for 
$\lambda$ describing a gas phase.  The liquid and gas phases can coexist in the 
range of $\lambda_{0}$ between the limit of 
existence of the gas phase, $\lambda_{0a}$, and that of liquid, 
$\lambda_{0b}$.

If the condition 
\begin{equation}
\label{condition}
p^{*}\exp[(2\lambda_{0}/p^{*1/2})\ln(1 + \pi/2)] \gg 1
\end{equation}
holds, then the explicit $\lambda_{0}$-dependence of the properties of the 
gas phase can be deduced from Eqs.(\ref{pqexpansionfinal}) and 
(\ref{pgammaeq3})
\begin{equation}
\label{lambdagas}
\lambda = \lambda_{0}/p^{*1/2}
\end{equation}
\begin{eqnarray}
\label{qgas}
Q& -& Q_{0} = - (1/2) \ln(p^{*}/2)
\nonumber\\
&+ & {1 \over \sqrt{p^{*}}}[2\lambda_{0}\ln(1 + {\pi \over2}) 
-{1\over \sqrt{2}} e^{-(\lambda_{0}/p^{*1/2})\ln(1 + {\pi \over2})}]
\end{eqnarray}
We note the range of applicability of these results is rather wide - small 
pressure and nonzero $\lambda_{0}$, large pressure and arbitrary 
$\lambda_{0}$, and arbitrary pressure and large $\lambda_{0}$.  
The pressure dependence of the reduced length per particle $Q$ (\ref{qgas}) 
is the equation of state of the Morse gas.  

As the pressure increases, the distance between the minimum and maximum of the
right-hand-side of (\ref{pgammaeq3}) decreases, and at the critical pressure 
$p^{*} = p^{*}_{c} \simeq 0.0185$ the difference between the properties of 
liquid and gas disappears for the first time.  At this pressure and 
$\lambda_{0c} \simeq 0.4387$ the size of the liquid-gas coexistence region
shrinks to a point.

The negative pressure analysis is similar to what we did for $p^{*} = 0$ in 
Section IVC.  For negative pressure of sufficiently small magnitude the 
right-hand-side of Eq.(\ref{pgammaeq3}) vanishes both at $\lambda = 0$ and 
$1 + 2p^{*}e^{2 \lambda \ln(1 + \pi/2)} = 0$ reaching a maximum in between.  If
De Boer's parameter $\lambda_{0}$ is below this maximum, then 
Eq.(\ref{pgammaeq3}) has two solutions for $\lambda$.  The smaller (physical) 
solution describes a metastable Luttinger liquid.  As $\lambda_{0}$ increases, 
the two solutions approach each other.  When $\lambda_{0}$ reaches the 
maximum of the right-hand-side of (\ref{pgammaeq3}), we are at the limit of 
mechanical stability of the system - no liquid can exist for larger 
$\lambda_{0}$.

Alternatively, for sufficiently small fixed $\lambda_{0}$ the height of the 
maximum of (\ref{pgammaeq3}) decreases upon increase of the magnitude of 
pressure, and at some $p^{*}_{ms}(\lambda_{0})$ the maximum of 
(\ref{pgammaeq3}) reaches the level of $\lambda_{0}$ thus bringing the system
to the limit of mechanical stability.  It is curious that in the
quantum case the
``liquid'' solution disappears {\it before} the condition 
$1 + 2p^{*}e^{2 \lambda \ln(1 + \pi/2)} = 0$ is reached.  Therefore at the 
stability threshold both the energy barrier (between ``broken'' ground-state 
and stretched metastable liquid) and sound velocity remain finite.  Only in 
the classical limit $\lambda_{0} \rightarrow 0$ do these quantities 
vanish.       

The pressure-De Boer's parameter diagram showing ranges of existence of liquid 
and gas is displayed in Fig.8.    
\begin{figure}[htbp]
\epsfxsize=4.0in
\vspace*{-0.5cm}
\hspace*{-0.75cm}
\epsfbox{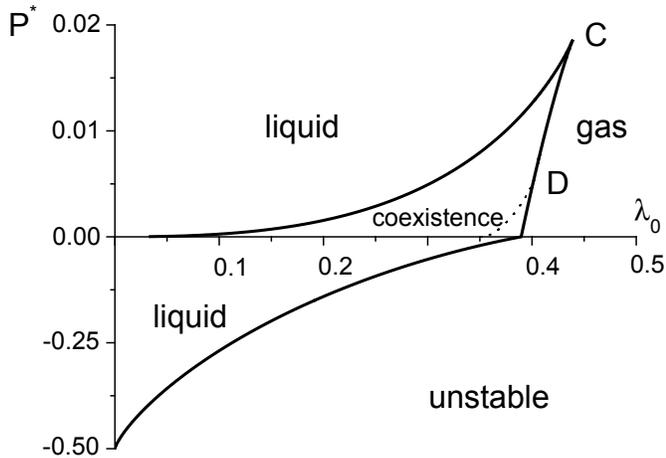}
\vspace*{-0.5cm}
\caption{The $p^{*}(\lambda_{0})$ diagram showing ranges of existence of 
liquid and monoatomic gas phases. Different scales are selected on the 
positive and negative parts of the pressure axis.  The liquid and gas can be 
in equilibrium along the dotted line.}
\end{figure}
We deliberately selected different scales on the positive and negative parts 
of the pressure axis in order to be able to show the complete picture.  As a
result of this choice there is an illusory change of slope of the line of 
mechanical stability of liquid at zero pressure - in reality the 
$p^{*}_{ms}(\lambda_{0})$ dependence is smooth.

The point C having coordinates $\lambda_{0c} \simeq 0.4387$, 
$p^{*}_{c} \simeq 0.0185$ where the limits of existence of liquid and gas meet 
is a candidate for the liquid-gas critical point.  Then the line of a 
liquid-gas evaporation transition should also pass through C.  This curve,
found by equating the ground-state energy (\ref{peofqfinal}) for both
phases is shown in Fig.8 by a dotted line.  It {\it does not} end at C, and 
everywhere within the DC segment of the metastability line the Luttinger liquid
has lower energy than the gas.  These results imply that a direct 
liquid-monoatomic gas transition is impossible.  Another gas phase,
diatomic, must intervene.  Although the existence of this phase will 
set phase boundaries at finite pressure, it will not affect the ranges of 
existence of the Luttinger liquid and monoatomic gas.
\begin{figure}[htbp]
\epsfxsize=3.79in
\vspace*{-0.5cm}
\hspace*{-0.75cm}
\epsfbox{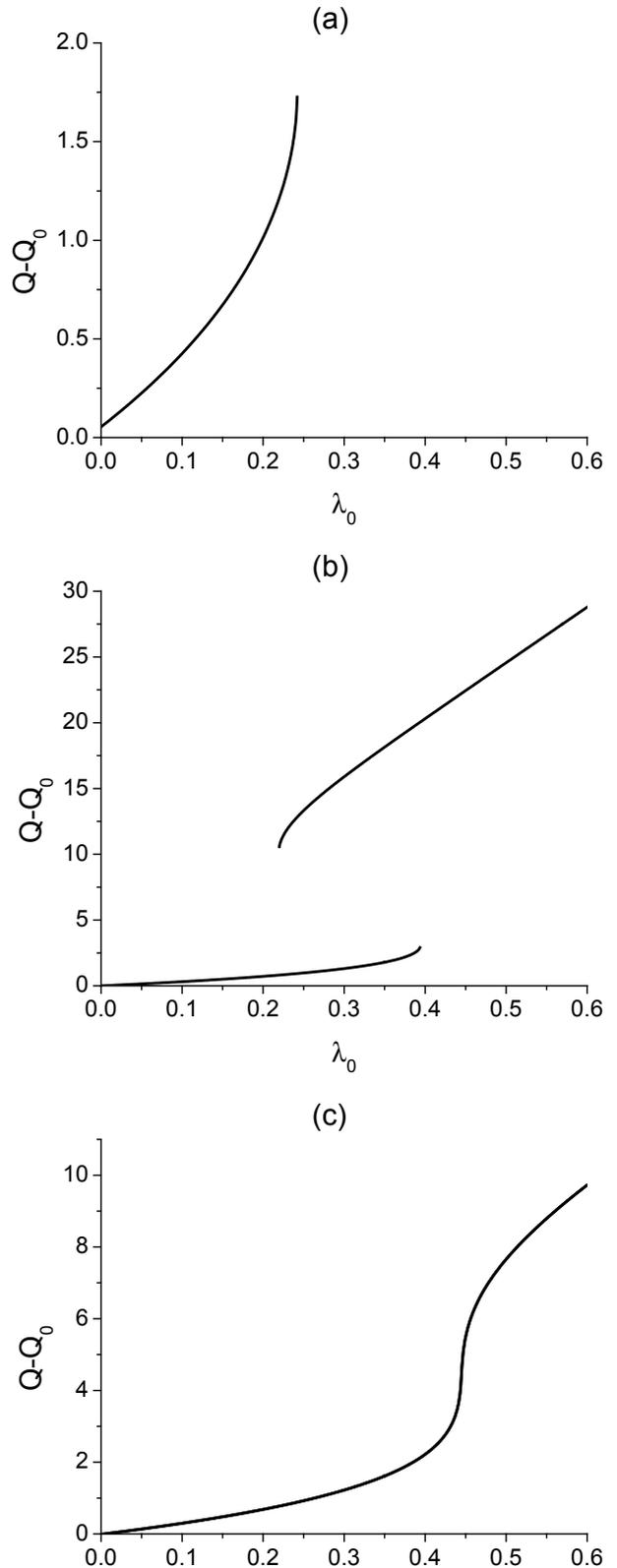}
\vspace*{-0.5cm}
\caption{The dependence of the reduced quantum expansion $Q - Q_{0}$ on De 
Boer's parameter $\lambda_{0}$ at a negative pressure of large magnitude (a), 
within the liquid-gas coexistence region (b), and past the point C of 
Fig.8 (c).}
\end{figure}
\noindent
Other finite pressure properties of the system can be readily found.  As an 
example, Fig.9 illustrates the dependence of the reduced quantum expansion 
$Q - Q_{0}$ on  De Boer's number $\lambda_{0}$ for three different pressures.  
The curve of Fig.9a shows the expansion of the Luttinger liquid due to
the combined effect of zero-point motion and negative pressure of 
sufficiently large magnitude ($p^{*}= -0.1$ was used).  The end point of the 
dependence is the limit of mechanical stability of the liquid.  In Fig.9b the
pressure is selected to be within the region of the liquid-gas coexistence 
($p^{*} = 0.002$ was used).  As a result the quantum expansion dependence is
not unique:  the lower (larger density) curve corresponds to the liquid while 
the upper (lower density) curve curve is for the monoatomic gas.  The
end of the ``liquid'' curve and the beginning of the ``gas'' curve are
limits of existence of these phases.  In Fig.9c the pressure is
selected to satisfy 
the condition $p^{*} > p^{*}_{c}$ i.e. past the point C of Fig.8.  Now there 
is no quantitative difference between liquid and gas.  The presence of a 
relatively steep part around $\lambda_{0} \simeq 0.44$ is the effect of the
proximity of $p^{*} = 0.02$ to $p^{*}_{c} \simeq 0.0185$.

To test the limits of applicability of our theory we also investigated the 
reduced sound velocity and relative bond fluctuation at various pressures.
We found that at fixed $\lambda_{0}$ the reduced sound velocity initially
increases with pressure but then at pressures exceeding the level of about
$6000$ or larger, it begins to decrease vanishing at the point where the bond 
length $Q$ vanishes.  The fall and vanishing of the sound velocity do not 
correspond to physical reality and have their origin in inadequacy of the 
Morse potential approximation at small interparticle distances.  Similarly, the
relative fluctuation diverges upon approaching the nonphysical $Q = 0$ point.  
These artifacts do not pose practical limitations to our theory because they 
occur at unrealistically large pressures.   We verified that if we limit 
ourselves to pressures not exceeding $5000$, then the relative fluctuation 
is smaller than $0.5$, and the behavior of other properties of the system is 
in agreement with physical expectations.
\begin{table}
\caption{Morse parameters for a series of molecular substances and some of 
their computed properties at zero pressure.  Substances are arranged in the 
order of decrease of their De Boer's number $\lambda_{0}$.  Blank entries 
correspond to gas ground states when interparticle separation is infinite.}

\begin{tabular}{ccccccccc}
&
\( m \)&
\( \epsilon  \)&
\( \ell  \)&
\( H_{0} \)&
\( \lambda _{0} \)&
\( Q-Q_{0} \)&
\( Q \)&
\(-E{*} \)\\
&
a.u.&
K&
\AA&
\AA&
&
&
&
$10^{-3}$\\
\hline
\( H\uparrow  \)&
1.008&
6.19&
.6869&
4.153&
.9136&
&
&
$0$\\
\( D\uparrow  \)&
2.014&
6.19&
.6869&
4.153&
.6463&
&
&
$0$\\
\( T\uparrow  \)&
3.016&
6.19&
.6869&
4.153&
.5281&
&
&
$0$\\
\( ^{3}He \)&
3.016&
10.8&
.5350&
2.980&
.5134&
&
&
$0$\\
\( ^{4}He \)&
4.003&
10.8&
.5350&
2.980&
.4456&
&
&
.0509\\
\( H_{2} \)&
2.016&
32.2&
.6900&
3.440&
.2819&
1.1860&
6.172&
159.6\\
\( D_{2} \)&
4.028&
32.2&
.6900&
3.440&
.1994&
.7178&
5.703&
376.6\\
Ne&
20.18&
35.6&
.5200&
3.110&
.1124&
.3591&
6.340&
632.6\\
Ar&
39.95&
120&
.6600&
3.860&
.0343&
.1005&
5.949&
883.9\\
\end{tabular}
\end{table}

\section{Applications and discussion}  

Fitting two-body potentials of molecular substances into the Morse form 
can provide us with the depth of the potential well $\epsilon$, the 
interaction range $l$, and the position of the potential minimum $H_{0}$.  
Supplemented by the masses of the underlying particles, this information 
is an input of our theory which then allows to determine De Boer's quantum 
parameter $\lambda_{0}$ (\ref{lambda_{0}}), the ground state and virtually any 
property.  

The fitting procedure can introduce uncertainties because real two-body 
interactions do not have the Morse form.  We already know that the Morse 
potential underestimates the strength of the overlap repulsion at short 
distances but as long as the reduced pressure does not exceed $5000$, this 
flaw is practically irrelevant.

The Morse potential also underestimates the magnitude of Van der Waals 
attraction at large distances.  We found that in the condensed state this  
shortcoming can be kept under the control by carrying out a Morse potential 
fit in a range of interparticle distances followed by a consistency check 
verifying that the segment of most probable particle location (formed by 
computed equilibrium bond length plus/minus its rms fluctuation) is well 
inside the fitting range.

The first four columns of Table I represent Morse parameters of various 
molecular substances which are used to compute De Boer's number $\lambda_{0}$
shown in the fifth column.  The remaining three columns are reduced quantum
expansion $Q - Q_{0}$, bond length $Q$, and the energy per particle $E^{*}$ 
(all at zero pressure) calculated using the theory developed in this paper.
The loci of all these substances are also indicated on the reduced energy 
curve $E^{*}(\lambda_{0})$ of Fig.6.  For the substances whose ground state 
is the Luttinger liquid, Fig.5a also shows the magnitudes of the reduced 
classical bond length $Q_{0}$ and its relative fluctuation $f^{*}/Q$, while 
Fig.5b gives the values of the Luttinger liquid exponent $g$.  

In computing these properties we also assumed that the three-dimensional form 
of the interaction does not change upon one-dimensional confinement of the 
particles and that translational symmetry is preserved.   Both these 
assumptions are approximations if the confinement is achieved in carbon 
nanotube bundles because interparticle interaction is mediated by
the carbon environment \cite{Kostov} while the axial motion takes place in a 
periodic potential \cite{dmc1}.  The former effect generally weakens 
interparticle attraction at large distances thus making the system more 
quantum.  On the other hand the external periodic potential due to the carbon
environment has an opposite effect leading to upward renormalization of the
mass.  Therefore the properties of strictly one-dimensional matter may  
differ qualitatively from those of the matter inside nanotube bundles.  The 
effect of an axial periodic potential may be even more dramatic, and the 
effective mass approximation insufficient if the corrugation is strong
enough to introduce a commensurate-incommensurate phase transition 
\cite{cic}.  This potentially important effect is beyond the scope of
our theory and cannot be discussed here.

Before considering individual substances it is useful to look at the 
properties of one-dimensional matter as a whole and compare them with those
of laboratory substances.  The corresponding states analysis of 
three-dimensional molecular matter is  based on the Lennard-Jones pair 
potential and except for the vicinity of the classical limit 
$\lambda_{0} = 0$, it is empirical \cite{deBoer,Anderson&Palmer,Clark&Chao}.  

The substances in Table I are arranged in the order of decrease of their 
De Boer's number $\lambda_{0}$ which is naturally the same as in three 
dimensions.  The main qualitative difference from the ordinary substances
occurs because of the dominant role played by zero-point motion which in one
dimension forbids the crystal ground state.  We find that spin-polarized 
isotopes of hydrogen (hydrogen $H\uparrow$, deuterium $D\uparrow$, and 
tritium $T\uparrow$) and $^{3}He$ are monoatomic gases, $^{4}He$ is diatomic 
gas, while molecular hydrogen and heavier substances are Luttinger liquids.  
If we view the Luttinger liquid as the counterpart of the crystal in three 
dimensions, then the bold part of the reduced energy curve, Fig.6, closely 
resembles its three-dimensional counterpart \cite{Anderson&Palmer}.  There is
a change of slope somewhere between $H_{2}$ and $^{4}He$ which in our case is 
dissociation of the Luttinger liquid into a diatomic gas while in the 
three-dimensional world it is a melting transition.

Our dependence of the reduced Debye temperature $\theta^{*}$ on De Boer's 
number $\lambda_{0}$, Fig. 5b, also looks very similar to its 
three-dimensional counterpart \cite{deBoer}, and even empirical 
values of the reduced Debye temperature are close 
to their computed one-dimensional analogs.  We also find that 
$\theta^{*}(\lambda_{0})$ dependence has a maximum somewhere past molecular 
hydrogen;  from empirical data it seems impossible to tell whether this effect 
is present or not in three dimensions.

Before comparing quantum expansion in one and three dimensions, we note that
the quantum theorem of the corresponding states applied to the Lennard-Jones 
system predicts that the reduced volume per particle (length in one dimension) 
$Q$ is only determined by De Boer's quantum parameter $\lambda_{0}$.  At the 
same time for the Morse system the analogous statement is valid for the 
reduced quantum expansion $Q - Q_{0}$. However the inspection of Table I 
(see also Fig. 5b) shows that the values of the reduced classical bond length 
$Q_{0} = H_{0}/l$ belong to the relatively narrow interval roughly between 
$5$ and $6.3$.  Therefore the variation of $Q_{0}$ from substance to 
substance can be ignored and within experimental error our results can be 
compared to their empirical counterparts \cite{deBoer,Anderson&Palmer}.  Again
we find that the $\lambda_{0}$ dependencies of the reduced quantum expansion in
one and three dimensions are qualitatively similar.

In three dimensions all these properties can be computed perturbatively in the 
$\lambda_{0} \rightarrow 0$ limit \cite{deBoer,Anderson&Palmer} with the 
conclusion that to leading order 
$E^{*}(\lambda_{0}) - E^{*}(0)$, $Q - Q_{0}$, and $\theta^{*}$ all vanish 
linearly with $\lambda_{0}$.  This behavior is identical to the 
$\lambda_{0} \rightarrow 0$ limit of our theory.

As a final comment, we note that from the viewpoint of their electron 
transport properties all the molecular substances are normally insulators as 
they have completely filled electronic shells.  However at sufficiently large 
pressure when electron wave functions of neighboring molecules overlap 
considerably, any substance should turn into a metal \cite{Wigner}.  At that 
point our ``molecular'' approximation describing many-body physics in terms 
of additive two-body interactions fails.  Typically this happens at a very 
large pressure, and a different approach explicitly accounting for the 
dynamics of the electron degrees of freedom is necessary.  This complex 
problem is beyond the scope of the present paper.  For the case of molecular 
hydrogen, however, we will be able to estimate this critical pressure when 
the metal-insulator transition takes place without leaving the framework of 
our theory.  

In subsequent discussion of individual substances we first present the results 
based on the pair interaction potential in free space.  These conclusions are 
robust.  On the other hand, our comments about the properties of matter inside 
nanotubes are speculative as they rely on the assumption that the effect
of carbon environment can be accommodated within the framework of our theory by
adjusting De Boer's parameter $\lambda_{0}$.  It is also important to keep in 
mind that inside nanotube bundles there will be an additional interaction 
between different one-dimensional channels filled with absorbed substances.  
This interaction is responsible for exotic crossover effects which can be 
viewed as an effective change of space dimensionality \cite{Calbi}.  These 
effects are also beyond the scope of our one-dimensional theory.  

\subsection{Spin-polarized hydrogen and its isotopes}

The pair interaction between two particles of the spin-polarized hydrogen
family has been computed by Kolos and Wolniewicz \cite{KW}.  Etters, Dugan 
and Palmer \cite{EDP} have found a very good Morse fit to the 
Kolos-Wolniewicz potential;  the Morse parameters shown in Table I are their 
values.   

Compared to the other elements in Table I, these substances have the 
shallowest potential well which is only $6.19 K$ deep. Combined with 
its smallest mass, this makes $H\uparrow$ the ``quantummost'' element with 
$\lambda_{0} = 0.9136$.  Spin-polarized deuterium $D\uparrow$ is second in
line with $\lambda_{0} = 0.6463$ while spin-polarized tritium $T\uparrow$ 
takes the third place, $\lambda_{0} = 0.5281$.  All these elements are 
monoatomic gases at zero pressure as can be seen from Fig.6.  On the other 
hand at zero pressure in three dimensions the heaviest of the family, 
$T\uparrow$, forms a liquid while $H\uparrow$ and $D\uparrow$ are gases 
\cite{EDP}.

External pressure confines these gases to a finite density but because De 
Boer's numbers are larger than $\lambda_{0c} \simeq 0.4387$ corresponding to 
point C of Fig.8, applying pressure is not going to turn them into liquids.

\subsection{Helium}

The pair interaction between two helium atoms is accurately described by the
semi-empirical Aziz potential \cite{Aziz} which is $10.8K$ deep;  this is the
second entry in Table I.  The authors of Ref.\cite{direct} proposed the 
Morse fit of the Aziz potential with the parameters $l = 0.5828 \AA$ and 
$H_{0} = 2.89\AA$ claiming that ``the integrated square of the deviation of 
the fit from the Aziz potential does not exceed $1\%$ in the range of 
localization of a $He$ atom''.  These parameters produce 
$\lambda_{0} \simeq 0.41$ for $^{4}He$ which according to our theory makes 
it a diatomic gas.  It is indeed experimentally known \cite{hedimer} that 
$^{4}He$ can form very large dimers with the bond length of $52\AA$.  We
verified however that at interparticle distances that large the Morse fit 
proposed in Ref.\cite{direct} is very poor.

Our own attempts to improve the fit increased the value of $\lambda_{0}$ 
bringing it into a narrow vicinity of the dimer dissociation threshold 
$\lambda_{02} = \sqrt2/\pi$, and without extra knowledge we could not make a 
decision whether $\lambda_{0}$ is larger or smaller than $\lambda_{02}$.  We 
resolved this dilemma by invoking the experimental result \cite{hedimer} that 
the binding energy of the $^{4}He$ dimer is $-1.1mK$. Halving this 
value and dividing the outcome by $10.8K$, the depth of the $He-He$
potential well, produces the last entry in Table I, the reduced energy per 
particle of the diatomic gas.  This can be substituted into Eq.(\ref{dimer}) 
to recover the fifth entry, $\lambda_{0} = 0.4456$.  As expected, 
this is only marginally smaller than the dimer dissociation threshold 
$\lambda_{02}$. Using the definition of De Boer's number, 
Eq.(\ref{lambda_{0}}), we can now recover the interaction range (the third 
entry in Table I) to be $l = 0.5350\AA$.  Finally the position of the minimum 
of the Morse potential $H_{0} = 2.980\AA$ was chosen to optimize the fit.

Upon application of pressure the gas of $^{4}He$ dimers will turn into a 
Luttinger liquid;  calculation of the properties of this dimer liquid is 
beyond the scope of our method.

De Boer's quantum parameter for $^{3}He$ can be obtained from that for 
$^{4}He$ by invoking their mass ratio.  This gives us the value quoted
in the fifth column of the $^{3}He$ row in Table I.   It is higher than the 
dimer dissociation threshold $\lambda_{02}$ thus ruling out earlier 
prediction \cite{ebashkin} that $^{3}He$ can form a dimer in one dimension.  
A many-body system of $^{3}He$ particles in one dimension will form a 
monoatomic gas with properties close to those of spin-polarized
tritium.  Similar to $T\uparrow$, the $^{3}He$ gas will not condense
under pressure.
    
Our result that $^{4}He$ forms a diatomic gas strictly in one dimension is in 
variance with earlier work \cite{dmc1,dmc2,krot} which predicted a liquid
ground state with the binding energy of order a few to tens $mK$.  This is the
same order of magnitude as the energy per particle in the diatomic gas.  
However Refs.\cite{dmc2,krot} also predict a liquid-solid phase transition 
which is forbidden in one dimension.

In applying our results to nanotubes one has to bear in mind that $^{4}He$ 
atoms are strongly attracted to the interstitial channels inside nanotube 
bundles.  The corrugation felt by the individual atom is so strong that the
effective mass enhancement is very large: $m^{*} \simeq 18m$ \cite{ebner}.  
There is also a weaker opposing effect: $28\%$ reduction in the
well depth of the pair interaction mediated by the carbon environment 
\cite{Kostov}.  Combining these effects and assuming the interaction range 
does not change significantly, we find that De Boer's number will 
{\it decrease} by a factor of $3.6$  away from its purely one-dimensional 
value thus implying a liquid ground state.  Similar outcome is expected for 
$^{3}He$.         

\subsection{Molecular hydrogen}

The pair interaction between two hydrogen (or deuterium) molecules is commonly
described by the semi-empirical Silvera-Goldman potential \cite{SG}.  Fig.1 
shows this potential together with its Morse fit; the calculated Morse 
parameters are quoted in Table I.

As can be deduced from Figs.5 and 6, and Table I the many-body system of 
$H_{2}$ molecules is a Luttinger liquid with strongest effects of zero-point 
motion.  It is characterized by $-5.14K$ cohesive energy (ground-state energy
per particle) which is an $84\%$ reduction in magnitude away from the depth of 
the $H_{2}-H_{2}$ potential, largest Debye temperature of $121K$, largest 
Luttinger liquid exponent $g \simeq 0.1$, largest quantum expansion of 
$0.82\AA$, and largest, just under $18\%$, relative fluctuation of the bond 
length.  

The equilibrium distance between the $H_{2}$ molecules is $4.26\AA$.  As can 
be seen from Fig.1 in a range around the equilibrium bond length significantly 
exceeding its rms fluctuation the Morse potential is a very good approximation
to the Silvera-Goldman potential.  The fit worsens at interparticle separations
exceeding $5\AA$; there are also deviations from the Silvera-Goldman potential
at distances smaller than $3\AA$.  However at these compressions Hemley and 
collaborators \cite{Hemley} have found a softening effect unaccounted for by
the Sivera-Goldman potential.  We verified that the Morse potential shown in 
Fig.1 provides a very good fit to the Hemley-corrected version of the 
Silvera-Goldman potential.       

Previous work \cite{dmc3} finds a liquid ground-state with the energy per 
particle to be $-4.8K$ and the bond length of $4.6\AA$.  These values are
close to our results.  However we disagree with the existence of a 
liquid-solid transition found in Ref.\cite{dmc3} at higher density; such a
transition is forbidden in one dimension.  

The building blocks of one-dimensional molecular hydrogen are $H_{2}$ molecules
whose size of $0.75\AA$ \cite{Wigner} is significantly smaller than the 
computed intermolecular distance of $4.26\AA$.  Such structure can be 
understood qualitatively from a complementary viewpoint:

There is exactly one electron per every hydrogen atom, and if the protons are
arranged equidistantly, then the valence band is half-full, and the resulting 
system is an alkali metal \cite{Ashcroft&Mermin}.  However 
Peierls \cite{Peierls} noticed that in one dimension the energy can be further
lowered by displacing every second nucleus by a prescribed distance. As a 
result of the period doubling the valence band becomes full, and the resulting 
dimer chain is an insulator.  We conclude that one-dimensional molecular 
hydrogen is an example of Peierls-distorted one-dimensional structure;  the 
dimers are hydrogen molecules.  

Peierls' arguments rely on an adiabatic approximation which ignores zero-point 
motion of the nuclei;  the former may change the answer qualitatively.  Our
theory which starts from interacting $H_{2}$ molecules shows that even for 
one-dimensional hydrogen the ground-state is a Peierls-distorted insulator 
despite strong quantum fluctuations.

This conclusion may change upon application of pressure which brings hydrogen
molecules closer to each other and increases relative fluctuation of the bond
length.  We argue that the distortion disappears and thus an insulator-metal
transition takes place when all the hydrogen atoms become translationally 
identical.  This transition is a one-dimensional version of the 
metallization transition predicted by Wigner and Huntington \cite{Wigner}.  
In one dimension the mechanism of the transition consists in ``undoing''  
the Peierls distortion.  

Since the bond between the two hydrogen molecules is significantly 
softer than that holding the $H_{2}$ molecule together, we assume that upon 
application of pressure, only the former decreases.  Thus the translational
equivalence of all the hydrogen atoms will be achieved when intermolecular 
spacing reaches the value of order $1.5\AA$, twice the size of the 
$H_{2}$ molecule.  With zero-pressure intermolecular spacing being 
$4.26\AA$, this corresponds to compression by a factor of 2.84.  The 
corresponding reduced pressure $p^{*} \gg 1$ can be found by inverting 
Eq.(\ref{qgas}):
\begin{equation}
\label{metal}
p^{*} \simeq 2e^{2\{Q_{0} - Q 
+ [\sqrt{2}\lambda_{0}\ln(1 + {\pi \over2}) - {1 \over2}]e^{Q - Q_{0}}\}}
\end{equation} 
Substituting here $Q_{0} = 4.986$, $Q = H/l = 2.174$, and 
$\lambda_{0} = 0.2819$ we arrive at the reduced pressure of $545$.  Our theory 
which does not explicitly consider electronic degrees of freedom fails in the 
vicinity of the inverse Peierls transition.

For molecular hydrogen in one dimension the unit of pressure is a {\it force} 
of $\epsilon/l \simeq 6.44* 10^{-12}N$ strong.  Multiplying this by $545$ 
we find that $3.51 * 10^{-9}N$ force compressing one-dimensional 
hydrogen may suffice to induce a transition into metallic state.  If this 
force is applied at the $1\AA^{2}$ area, then the corresponding 
three-dimensional pressure will be $351GPa$.    We note that three-dimensional
solid hydrogen subject to pressure that big still resists metallization 
\cite{pressure}.  Unfortunately the accuracy of our estimate is not great 
because of the exponential dependencies in (\ref{metal}) - the actual 
one-dimensional transition may happen at lower or larger pressures.

For hydrogen confined inside interstitial channels of carbon nanotube bundles 
the carbon environment effectively reduces the well depth of the pair 
interaction by $54\%$ \cite{Kostov}.  This effect alone would suffice to 
turn the many-body system of hydrogen molecules into a gas of $(H_{2})_{2}$ 
complexes.  However if the effective mass enhancement is comparable to that for
$He$ \cite{ebner}, the liquid ground state might be restored.

As can be seen from Fig.6 and Table I, the ground state of 
molecular deuterium is a Luttinger liquid.  The quantitative difference from 
the properties of molecular hydrogen is solely due to the fact that $D_{2}$
has a larger mass. The cohesive energy of the one-dimensional $D_{2}$ liquid is
$-12.1K$ ($62\%$ reduction in the magnitude of the $D_{2}-D_{2}$ pair 
potential well), the Debye temperature is $99.8K$ while the equilibrium 
distance between $D_{2}$ molecules is $3.94\AA$.

\subsection{Heavier substances}

While discussing the physics of molecular hydrogen in one dimension we came
to the conclusion that it can be viewed as an example of Peierls-distorted 
structure.  The same arguments are applicable to any element with odd number of
electrons:  the period doubling should take place and the resulting system must
be an insulator.  We note that three-dimensional counterparts of these 
substances are metals.  These observations imply that our theory is also 
applicable to substances which traditionally are not considered to belong to
the molecular group.   For example, one-dimensional lithium must be an 
insulating Luttinger liquid of $Li_{2}$ molecules \cite{Li}.  Similar to 
molecular hydrogen, under pressure it should undergo a metallization 
transition.  Had we known the pair interaction between two $Li_{2}$ molecules, 
we could have computed the properties of the lithium liquid.  From the 
viewpoint of the quantum theorem of corresponding states molecular lithium is 
expected to occupy a place somewhere between $D_{2}$ and $Ne$.         

Table I also contains the Morse data for $Ne$ and $Ar$.  They were obtained
from the parameters of the Lennard-Jones interaction potential 
\cite{Ashcroft&Mermin} which is commonly used to describe these noble gases.  
Some of the properties of these substances in one dimension can be found in
Table I and in Figs.5 and 6.  The equilibrium interparticle spacing, 
cohesive energy and Debye temperature can be extracted from what is shown
in the same manner as was done for lighter elements.  As the underlying 
particles become heavier, the effect of zero-point motion decreases.  For 
elements heavier than $Ar$ quantum fluctuations can be ignored for most 
practical purposes.         

\section{ACKNOWLEDGMENTS}  

We thank  M. Fowler, I. Harrison R. Kalas and T. J. Newman for
valuable discussions.  

This work was supported by the Thomas F. Jeffress and Kate Miller Jeffress 
Memorial Trust, and by the Chemical Sciences, Geosciences and 
Biosciences Division, Office of Basic Energy Sciences, Office of Science, U. S.
Department of Energy.


\begin{references}
\bibitem{note1}  The first term of (\ref{waction}) written in terms of the 
Fourier transform $h(\omega)$  is identical to the first term of Eq.(5) 
in \cite{KS}.
\end{references}
\end{document}